\documentclass[10pt,superscriptaddress,aps,twocolumn,longbibliography,nofootinbib]{revtex4-2}

\usepackage{graphicx,color,graphpap}      
\usepackage{amsmath}
\usepackage{booktabs}
\usepackage{enumitem}
\usepackage{amssymb}
\usepackage{amsthm}
\usepackage[braket, qm]{qcircuit}
\usepackage{pstricks}
\usepackage{float}
\usepackage[colorlinks=true, citecolor=blue, linkcolor=red]{hyperref}
\usepackage[T1]{fontenc}
\usepackage{bbm}
\usepackage{dsfont}
\usepackage[linesnumbered,ruled,vlined]{algorithm2e}
\SetKwInput{kwInit}{Init}
\usepackage{mathtools}

\usepackage{tikz}
\usetikzlibrary{positioning}
\usetikzlibrary{matrix}
\usepackage{subcaption,graphicx}
\usepackage[ruled,vlined]{algorithm2e}
\usepackage{siunitx}
\usepackage{qcircuit}
\usepackage{pifont} 
\usepackage{listings}
\usepackage{multirow}
\usepackage{xcolor}


\definecolor{codegreen}{rgb}{0,0.6,0}
\definecolor{codegray}{rgb}{0.5,0.5,0.5}
\definecolor{codepurple}{rgb}{0.58,0,0.82}
\definecolor{backcolour}{rgb}{0.95,0.95,0.92}

\lstdefinestyle{mystyle}{
    backgroundcolor=\color{backcolour},
    commentstyle=\color{codegreen},
    keywordstyle=\color{magenta},
    numberstyle=\tiny\color{codegray},
    stringstyle=\color{codepurple},
    basicstyle=\ttfamily\small,
    breaklines=true,
    numbers=left,
}
\newcommand{\secref}[1]{Sec.~\ref{#1}}

\newcommand{\figref}[1]{Fig.~\ref{#1}}

\renewcommand{\eqref}[1]{Eq.~(\ref{#1})}

\newtheoremstyle{example}{\topsep}{\topsep}%
{}
{}
{\bfseries}
{:}
{   }
{\thmname{#1}\thmnumber{ #2}}
\theoremstyle{example}

\theoremstyle{definition}

\newtheorem*{theorem*}{Theorem}

\def\orcid#1{\kern -0.4em\href{https://orcid.org/#1}{\includegraphics[keepaspectratio,width=0.7em]{orcid_logo.pdf}}}

\usepackage{lipsum}

\long\def\ca#1\cb{} 
\begin{document}
\title{Testing platform-independent quantum error mitigation on noisy quantum computers}

\author{Vincent Russo}
\thanks{Corresponding author: \href{mailto:vincent@unitary.fund}{vincent@unitary.fund}.}
\affiliation{Unitary Fund}

\author{Andrea Mari}
\affiliation{Unitary Fund}

\author{Nathan Shammah}
\affiliation{Unitary Fund}

\author{Ryan LaRose}
\affiliation{Unitary Fund}
\affiliation{Institute of Physics, \'{E}cole Polytechnique F\'{e}d\'{e}rale de Lausanne (EPFL), CH-1015 Lausanne, Switzerland}

\author{William J. Zeng}
\affiliation{Unitary Fund}
\affiliation{Goldman Sachs \& Co., New York, NY}

\begin{abstract}
    We apply quantum error mitigation techniques to a variety of benchmark problems and quantum computers to evaluate the performance of quantum error mitigation in practice. To do so, we define an empirically motivated, resource-normalized metric of the improvement of error mitigation which we call the improvement factor, and calculate this metric for each experiment we perform. The experiments we perform consist of zero-noise extrapolation and probabilistic error cancellation applied to two benchmark problems run on IBM, IonQ, and Rigetti quantum computers, as well as noisy quantum computer simulators. Our results show that error mitigation is on average more beneficial than no error mitigation --- even when normalized by the additional resources used --- but also emphasize that the performance of quantum error mitigation depends on the underlying computer.
\end{abstract}

\maketitle


\section{Introduction}\label{sec:intro}

Quantum computers have steadily improved over the past two decades as can be seen in component metrics like T1 and T2 times~\cite{huang2020superconducting, sevilla2020forecasting} as well as full system metrics like quantum volume~\cite{cross2019validating}. While we expect these hardware improvements to continue, it is generally accepted that error rates cannot be made low enough purely by hardware improvements. Rather, to achieve error rates low enough for useful applications, hardware improvements should be coupled with algorithmic or software methods.

The most commonly pursued algorithmic method is quantum error correction~\cite{shor1995scheme, calderbank1996good, steane1996error}, which generally provides a tradeoff in qubit quantity for qubit quality --- i.e., using more qubits to achieve a lower logical error rate. Today, the state-of-the-art experiments in quantum error correction~\cite{chen2021exponential,acharya2022suppressing} confirm an exponential suppression of errors as the code distance increases but do so for relatively small code distances. For example, the largest surface code implementation to our knowledge~\cite{acharya2022suppressing} uses $49$ physical qubits to encode one logical qubit in a distance five surface code, while rough estimates for current error rates require around $1000$ physical qubits per logical qubit for fault tolerance. 

Because the experimental requirements of quantum error correction are very demanding, and because of widespread interest in applications of noisy quantum computers~\cite{preskill2018quantum}, a new set of algorithmic methods to deal with errors has emerged in recent years. These new methods are referred to as quantum error mitigation~\cite{endo2021hybrid, cai2022quantum} and are designed to be less experimentally demanding than full quantum error correction. However, this comes at the cost of being less general and more heuristic than quantum error correction. 


While a relatively large number of error mitigation techniques have been proposed~\cite{song2019quantum, vuillot2017error, kandala2019error, giurgica2020digital, urbanek2020error, google2020hartree, zhang2020error}, there have been relatively few experiments using error mitigation, despite the fact that error mitigation is specifically designed for current quantum computers. A summary from the literature of quantum error mitigation experiments performed on quantum computers is shown in Table~\ref{tab:qem-experiments}. Note that this table is not exhaustive for all benchmarks outside of the context of quantum error mitigation. For instance, this review~\cite{stano2022review} provides a thorough collection of T1 and T2 times for the dynamical decoupling benchmark.  

\begin{table*}
    \centering
    \begin{tabular}{|c|c|c|c|c|} \hline 
        QEM                  & Benchmark & Qubits $n$       & Computer(s)                                   & Ref.                    \\ \hline
        \multirow{8}{*}{ZNE} & RB        & $1$, $2$ & $5$-qubit superconducting device              & \cite{kandala2019error} \\
                             & RB        & $2$       & IBMQ London \& Rigetti Aspen-8                & \cite{larose2020mitiq}  \\ 
                             & RB, PG    & $2$ - $5$ & IBMQ Lagos \& IBMQ Casablanca                 & \cite{cirstoiu2022volumetric} \\       
                             & RB, MC    & $3, 5, 12$      & IBMQ Lima, IBMQ Kolkata, Rigetti Aspen-M2, IonQ Harmony
                             & (This work)  \\                      
                             & VQE       & $4$       & $5$-qubit superconducting device              & \cite{kandala2019error} \\ 
                             & QV        & $5$       & IBMQ Belem, IBMQ Lima, \& IBMQ Quito          & \cite{larose2022error}  \\ 
                             & RB, TE    & $26$      & $27$-qubit superconducting device             & \cite{kim2021scalable}  \\ 
                             \hline 

        \multirow{2}{*}{PEC} & RB        & $2$       & $2$-qubit trapped ion ($^{171}$Yb$^+$) device & \cite{zhang2020error} \\
                            & RB, MC     & $3$      & Rigetti Aspen-M2, IBMQ Lima, IonQ Harmony             & (This work) \\         
                             & CYC       & $4$      & $4$-qubit superconducting device               & \cite{ferracin2022efficiently} \\ 
                            & TE, CYC   & $4$, $10$ & $27$-qubit superconducting device             & \cite{berg2022probabilistic} \\     
                             \hline 

        \multirow{6}{*}{DD}  & IDLE      & $1$, $2$ & IBMQX4, IBMQX5, \& Rigetti Acorn              & \cite{pokharel2018demonstration} \\ 
                            & Adder, GHZ, QAOA, QFT, VQE      & $4$, $5$, $6$ & IBMQ Guadalupe, \& IBMQ Jakarta             &  \cite{smith2022timestitch} \\        
                             & QPE       & $5$       & IBMQ Paris, IBMQ Guadalupe, \& IBMQ Toronto   & \cite{das2021adapt} \\
                             & QV        & $6$       & IBMQ Montreal                                 & \cite{jurcevic2021demonstration} \\
                             & QFT       & $6$, $7$ & IBMQ Paris, IBMQ Guadalupe, \& IBMQ Toronto             & \cite{das2021adapt} \\  
                             & BV        & $7$, $8$ & IBMQ Paris, IBMQ Guadalupe, \& IBMQ Toronto             & \cite{das2021adapt} \\  
                             & QAOA      & $8$, $10$ & IBMQ Paris, IBMQ Guadalupe, \& IBMQ Toronto             &  \cite{das2021adapt} 
                             \\  \hline
        \multirow{2}{*}{CDR} & RB, PG        & $2$ - $5$ & IBMQ Lagos \& IBMQ Casablanca                 & \cite{cirstoiu2022volumetric} \\ 
                            & VQE        & $5$      & IBMQ Rome                                     & \cite{czarnik2021error}       \\ 
                            & VQE        & $6$      & IBMQ Toronto                                  & \cite{czarnik2022improving}   \\ 
                            & VQE        & $16$     & IBMQ Almaden                                  & \cite{zhang2021variational}   \\ \hline
        \multirow{2}{*}{SSE} & VQE       & $2$       & $2$-qubit superconducting device              & \cite{colless2018computation} \\ 
                             & VQE       & $2$       & $3$-qubit superconducting device              & \cite{sagastizabal2019experimental} \\ \hline 
        \multirow{1}{*}{VD} & GHZ       & $5$       &  $5$-qubit trapped ion, UMD, ($^{171}$Yb$^+$) device               & \cite{seif2022shadow} \\ \hline                              
        
    \end{tabular}
    \caption{A history of quantum error mitigation experiments on quantum computers in literature. Quantum error mitigation (QEM) Technique acronyms: ZNE = zero-noise extrapolation, PEC = probabilistic error cancellation, also referred to as quasi-probabilistic decomposition (QPD) by some authors, DD = dynamical decoupling, SSE = subspace expansion, VD = virtual distillation, CDR = Clifford data regression. Benchmark acronyms: RB = randomized benchmarking, VQE = variational quantum eigensolver, TE = time evolution, IDLE = allowing a state to idle (identity operation), CYC = alternate cycles of single-qubit layers and Clifford layers, GHZ = Greenberger–Horne–Zeilinger, MC = mirror circuits, Adder = Ripple Carry Adder.}
    \label{tab:qem-experiments}
\end{table*}

In this work, we evaluate quantum error mitigation in practice using a suite of experiments on various benchmarks and quantum computers. We consider two error mitigation techniques, two benchmark problems, and four quantum computers. To quantify the performance of quantum error mitigation (relative to no error mitigation), we define a natural metric that we call the \emph{improvement factor}. 

Our results show that quantum error mitigation improves the performance of noisy quantum computations in nearly all experiments we consider, even when normalized by the additional resources (namely, samples) used in the error mitigation techniques. Depending on the number of qubits, circuit depth, and particular computer in the experiment, our results show between a $1$x and $7$x improvement from quantum error mitigation. Further, the error mitigation we use is ``out-of-the-box'' in that it is not tailored to the benchmark problems or computers we consider. Because of this, we expect quantum error mitigation to be an essential component of NISQ and even error-corrected computations and offer perspective on these points.  

The rest of the paper is organized as follows. Section~\ref{sec:methods} describes our methods for assessing the performance of quantum error mitigation in practice. This includes our definition of the improvement factor (\secref{subsec:improvement-factor}), the error mitigation techniques (\secref{subsec:qem-techniques}), the benchmark problems (\secref{subsec:benchmark-problems}), and the quantum computers (\secref{subsec:computers}) used in our experiments. 
 We present the results of our experiments in~\secref{sec:results}, and we discuss them in the larger context of quantum error mitigation and quantum computation in~\secref{sec:discussion}.

\section{Methods}\label{sec:methods}

\begin{figure*}
    \centering
    \includegraphics[width=1.0\textwidth]{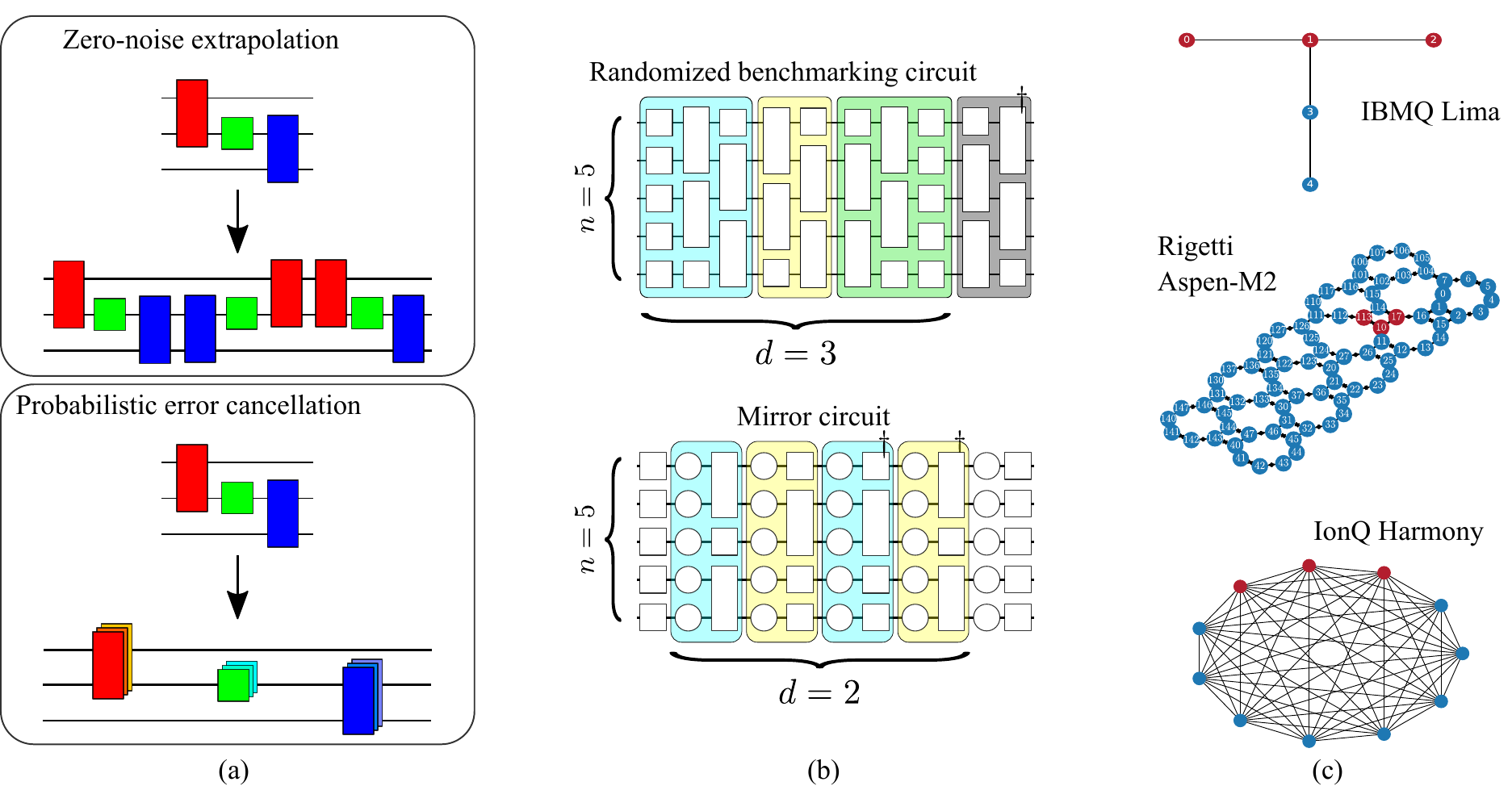}
    \caption{An overview of our method to assess the performance of quantum error mitigation in practice. An experiment consists of \textbf{(a)} a QEM technique, \textbf{(b)} a benchmark problem, and \textbf{(c)} a quantum computer. The result of an experiment is the improvement factor (Sec.~\ref{subsec:improvement-factor}) $\mu_{\rm QEM}$ defined in~\eqref{eq:if}, potentially at various numbers of qubits $n$ and/or circuit depths $d$. \textbf{(a)} Cartoon graphics of the two quantum error mitigation techniques we consider. In zero-noise extrapolation (Sec.~\ref{subsec:zne}), a circuit $C$ is mapped to a set of noise-scaled circuits by adding gates that compile (without noise) to the identity. The expectation value is computed for each noise-scaled circuit, and the results are extrapolated to zero-noise with either linear or Richardson extrapolation. In probabilistic error cancellation (Sec.~\ref{subsec:pec}), each ideal (unitary) gate of a circuit is expressed in the noisy basis of the computer. A number of new circuits are sampled from this expansion and executed results are combined to produce the error-mitigated result. \textbf{(b)} The two benchmark problems we use in our experiments (Sec.~\ref{subsec:benchmark-problems}). An $n$-qubit, depth $d$ mirror circuit $C$ is defined by a single layer of Clifford gates (white squares), $d$ Clifford layers followed by their inverses (rectangles in colored boxes, daggers denote inverses) with intermediate random Pauli gates (circles in colored boxes), and a final layer of Pauli (white circles) and Clifford gates. This sequence produces a single bitstring $|z_C\rangle$, and we take $\hat{A} = |z_C\rangle\langle z_C|$ as the observable. An $n$-qubit, depth $d$ randomized benchmarking circuit is defined by a random sequence of $d$  elements of the $n$-qubit Clifford group and a final inverse such that the final state is $|0\rangle$, and we take $\hat{A} = |0\rangle\langle 0|$ as the observable. \textbf{(c)} Qubit coupling maps for the four quantum computers we perform experiments on (see Sec.~\ref{subsec:computers} for device characteristics and error rates). Red nodes show qubits used for $n = 3$ qubit experiments. Qubit selection was not available on IonQ Harmony and so nodes are unlabeled. We also perform experiments on noisy quantum computer simulators (Sec.~\ref{subsubsec:noisy-simulators}).}
    \label{fig:methods-overview}
\end{figure*}

To assess the experimental performance of quantum error mitigation, we define a natural measure comparing the accuracy of an experiment with quantum error mitigation to the accuracy without quantum error mitigation. This measure, which we call the \textit{improvement factor}, is motivated and defined in Sec.~\ref{subsec:improvement-factor}. We experimentally calculate this measure using two quantum error mitigation techniques (Sec.~\ref{subsec:qem-techniques}) with two benchmark problems (Sec.~\ref{subsec:benchmark-problems}) on four quantum computers and three noisy quantum computer simulators (Sec.~\ref{subsec:computers}). All the error mitigation techniques for this study were implemented using the Mitiq error mitigating compiler~\cite{larose2020mitiq}.

\subsection{Improvement factor} \label{subsec:improvement-factor}

The goal of most quantum error mitigation techniques is to improve the estimation of expectation values. Following an empirical approach, we quantify the improvement of error mitigation by comparing the estimation errors obtained with and without error mitigation.

Let $\rho$ be an ideal $n$-qubit quantum state prepared by a noiseless quantum computer
after the execution of some given quantum circuit $C$, i.e., $\rho = C |0^{\otimes n} \rangle \langle 0^{\otimes n}| C^\dagger$. For an observable $\hat A = \hat{A}^\dagger$,
the ideal (noiseless) expectation value is
\begin{equation} \label{eqn:expectation-ideal}
    A = {\rm tr}[\rho \hat{A}] = {\rm tr}[ C |0^{\otimes n} \rangle \langle 0^{\otimes n}| C^\dagger \hat{A}].
\end{equation}
When using a noisy quantum computer, we instead prepare a noisy state $\rho'$ and collect $N$ shots (samples) to obtain an empirical estimate $A'$ of the expectation value.

%

The goal of quantum error mitigation (QEM) is to compute some quantity $A_{\rm QEM}$ which is a more accurate estimate of the ideal expectation value $A$ compared to the unmitigated estimate $A'$\footnote{Note that here and throughout, we use the acronym QEM to refer to a generic quantum error mitigation technique and a specific acronym for a specific quantum error mitigation technique. So, for example, the zero-noise extrapolated expectation value of $\hat{A}$ is denoted $A_{\rm ZNE}$, and similarly for other quantities. A summary of our notation is included in Appendix~\ref{app:notation}.}. Generally, computing $A_{\rm QEM}$ is done by executing a set of circuits $\{C_1, ..., C_{k_{\rm QEM} } \}$ related to $C$ --- usually with a different number of qubits, gates, and/or total shots $N_{\rm QEM}$ --- then post-processing the noisy results to obtain the error-mitigated estimate $ A_{\rm QEM}$.
%
%

We refer to an evaluation of an unmitigated expectation value $A'$ as a {trial}. After performing $t$ trials $A'_{[1]}, ..., A'_{[t]}$ we can quantify the estimation error through the root-mean-square error (RMSE) 
\begin{equation} \label{eqn:rmse}
   \sqrt{\frac{1}{t} \sum_{i=1}^{t} \left( A'_{[i]} - A \right)^2} .
\end{equation}
We use the RMSE since it reduces to the absolute error when all $A'_{[i]}$ are approximately equal (e.g. in the limit of large $N$) and, at the same time, it also takes into account the estimation error due to the statistical fluctuations of the results $A'_{[i]}$ over different trials.

Similarly, we refer to an evaluation of $A_{\rm QEM}$ as a {QEM trial}. After performing $t$ QEM trials $A_{\rm QEM}^{[1]}, ..., A_{\rm QEM}^{[t]}$ we evaluate the RMSE 
\begin{equation} \label{eqn:rmse-qem}
   \sqrt{\frac{1}{t} \sum_{i=1}^{t} \left(A_{\rm QEM}^{[i]} - A \right)^2} .
\end{equation}
As noted, evaluating each $A_{\rm QEM}^{[i]}$ potentially uses additional resources in the form of circuits, qubits, gates, and/or shots. To account for these additional resources, we define the \textit{problem-specific improvement factor}
\begin{equation} \label{eq:if-for-a-given-circuit-and-observable}
    \mu_{\rm QEM} (C, \hat{A}) := \frac{\sqrt{ N  \sum_{i=1}^{t} (A'_{[i]} - A )^2}}{\sqrt{N_{\rm QEM}  \sum_{i=1}^{t} (A_{\rm QEM}^{[i]} - A )^2}} , 
\end{equation}
i.e., the shot-normalized ratio of root-mean-square errors. (See Sec.~\ref{subsec:limitations-of-our-work} for a discussion on normalizing by other resources, e.g. qubits and gates, in addition to shots.) This value is a natural, empirically defined measure of the performance of quantum error mitigation for a specific expectation value problem defined by a circuit $C$ and observable $\hat A$. To generalize over different problems in addition to averaging over multiple trials, we also average over a set of circuits $\mathcal{C}$ and a set of observables $\mathcal{\hat{A}}$ to define the \textit{improvement factor}
\begin{equation} \label{eq:if}
    \mu_{\rm QEM} :=
    \frac{\sqrt{ N \sum_{C \in \mathcal{C}, \hat{A} \in \mathcal{\hat{A}}}
    \sum_{i=1}^{t} (A'_{[i]} - A )^2}}
    {\sqrt{N_{\rm QEM}\sum_{C \in \mathcal{C}, \hat{A} \in \mathcal{\hat{A}}}
    \sum_{i=1}^{t} (A_{\rm QEM}^{[i]} - A )^2}} .   
\end{equation}
Here, as in~\eqref{eq:if-for-a-given-circuit-and-observable}, the circuit $C$ is implicit in the expectation values $A$, $A'_{[i]}$, and $A_{\rm QEM}^{[i]}$, e.g. $A = {\rm tr} [ C |0\rangle \langle 0 | C^\dagger \hat{A}]$. While this definition is general with respect to the circuits $C$, experimentally we consider two classes of benchmark circuits (Sec.~\ref{subsec:benchmark-problems}) and quote the results from these classes of circuits separately. Indeed, for most experiments on quantum computers, we generate $\left| \mathcal{C} \right| = 4$ randomized instances of benchmark circuits from the two classes. We choose benchmark circuits such that there is one natural observable for each circuit, i.e. $\left| \mathcal{\hat{A}} \right| = 1$, where $|\cdot|$ denotes the cardinality of the set. In all cases, due to limited device availability, we perform $t = 1$ trial for each $C, \hat{A} \in \mathcal{C} \times \mathcal{\hat{A}}$.

We note that authors of~\cite{cirstoiu2022volumetric} also define a measure of the improvement from error mitigation, in particular a problem-specific measure. This quantity, which they call the relative mitigation error and denote by $\epsilon$, is given by (in the notation of this paper)
\begin{equation} \label{eq:epsilon}
    \epsilon_{\rm QEM} (C, \hat{A}) := \frac{| A_{\rm QEM} - A |}{|A' - A|}.
\end{equation}
For $t = 1$, $\mu_{\rm QEM} (C, \hat{A}) = \sqrt{\frac{N}{N_{\rm QEM}}} \epsilon^{-1}_{\rm QEM} (C, \hat{A})$.

\subsection{Quantum error mitigation techniques} \label{subsec:qem-techniques}


\subsubsection{Zero-noise extrapolation} \label{subsec:zne}

We apply zero-noise extrapolation (ZNE)~\cite{temme2017error, li2017efficient, kandala2019error} with both linear and Richardson extrapolation --- which we respectively denote ZNE(L) and ZNE(R) --- to our benchmark problems. For both cases, we evaluate $k_{\rm ZNE} = 3$ noisy expectation values $A'(\lambda_i)$ at different noise scale factors $\lambda_i \in \{1, 2, 3\}$ and the zero-noise limit is obtained as a linear combination of the results 
\begin{equation}\label{eq:zne}
   A_{\rm ZNE} = \sum_{i=1}^{k_{\rm ZNE}} \eta_i A'(\lambda_i) .
\end{equation}
For Richardson extrapolation, the best fit coefficients $\eta_i$ in \eqref{eq:zne} are given by~\cite{giurgica2020digital}
\begin{equation}
    \eta_i := \prod_{j \neq i} \frac{\lambda_j}{\lambda_j - \lambda_i}.
\end{equation}
For linear extrapolation, the coefficients $\eta_i$ are obtained from a linear best fit and also only depend on the noise scale factors, but the analytical expression is more involved (see Eq.~(26) of \cite{giurgica2020digital}).

For all ZNE experiments, we use global unitary folding~\cite{giurgica2020digital} to scale noise. For odd integer scale factors $\lambda_i$, this amounts to replacing the circuit $C$ by $C (C^\dagger C)^{(1 - \lambda_i)/2}$. If $\lambda_i$ is not an odd integer, a fraction of the full circuit is folded and appended to the circuit as described in ~\cite{giurgica2020digital}. Each noise-scaled circuit is executed with $\lfloor N / k_{\rm ZNE} \rfloor = \lfloor 10^4 / 3 \rfloor $ shots so that $N_{\rm ZNE} \simeq N = 10^4$ (i.e., so that we use the same total number of shots in ZNE as in the unmitigated experiment). For more details on our ZNE implementation, see Appendix~\ref{appx:zne}.

\subsubsection{Probabilistic error cancellation} \label{subsec:pec}

We also apply probabilistic error cancellation (PEC)~\cite{temme2017error, endo2018practical, zhang2020error} to each benchmark problem. Here, the first step is to characterize the set of noisy, implementable operations $\{\mathcal O_{\alpha}\}$ of a computer so that we can represent the ideal (noiseless) operations $\{ \mathcal G_i \}$ of a circuit in this basis, namely
\begin{equation}\label{eq:pec_rep}
    \mathcal G_i = \sum_\alpha \eta_{i, \alpha} \mathcal O_{\alpha} .
\end{equation}
Note that the calligraphic symbols $\mathcal G_i$ and $\mathcal O_{ \alpha}$ stand for super-operators acting on the quantum state of the qubits as linear quantum channels, and $\eta_{i, \alpha} \in \mathbb R$. In principle, this requires full tomographic knowledge of the noisy operations $\{ \mathcal O_{\alpha} \}$, but we make two simplifying assumptions in our experiments:
\begin{enumerate}
    \item We neglect errors of single-qubit gates.
    \item We assume that all two-qubit gates $\mathcal G_{2Q}$ (${\rm CNOT}$ or $\rm CZ$ in our experiments) are followed by local depolarizing noise, i.e.
        \begin{equation} \label{eq:cnot_noise_model}
            \mathcal G_{\rm 2Q}^{(\rm noisy)} = (\mathcal D_p \otimes \mathcal D_p) \circ  \mathcal G_{2Q},
        \end{equation}
        where $\mathcal D_p (\rho) = (1 - p) \rho + \frac{p}{3} ( X \rho X + Y \rho Y + Z \rho Z) $
        is the single-qubit depolarizing channel and $p$ is the local error probability.
\end{enumerate}
Under these assumptions, the quasi-probability representation of the ideal $ \mathcal G_{2Q}$ gate can be derived for any value of $p$~\cite{temme2017error, takagi2021optimal}. For each $ \mathcal G_{2Q}$ operation, we obtain $p$ from the error rate in the calibration data reported by the hardware vendor for the associated two-qubit gate (IBM, Rigetti), or from the average two-qubit error rate (IonQ). More precisely, in~\eqref{eq:cnot_noise_model} the overall two-qubit error probability is $p_{2Q} = 1 - (1-p)^2$. So, given the parameter $p_{2Q}$ reported by the calibration data of the computer, we estimate $p = 1 - \sqrt{1 - p_{2Q}}$ and use this in~\eqref{eq:cnot_noise_model} to obtain the basis $\mathcal O_{ \alpha}$.




After obtaining the basis $\mathcal O_{\alpha}$, we represent all two-qubit gates $G_i$ in the circuit in this basis as in~\eqref{eq:pec_rep} and stochastically sample $k_{\rm PEC} = 100$ new circuits to execute. Each circuit is executed with $N / k_{\rm PEC} = 10^4 / 100 $ shots so that $N_{\rm PEC} = N = 10^4$ (i.e., so that we use the same total number of shots in PEC as in the unmitigated experiment). For more details on our implementation of PEC, see Appendix~\ref{appx:pec}.

\subsection{Benchmark problems} \label{subsec:benchmark-problems}


A benchmark problem is defined by an $n$-qubit, depth $d$ quantum circuit $C$, and an observable $\hat A$ as in~\eqref{eqn:expectation-ideal}. In this work, we consider two benchmark problems in which the circuit $C$ produces (without noise) a single bitstring $z_C \in \{0, 1\}^n$, and we always take $\hat{A} = |z_C \rangle \langle z_C|$ as the corresponding observable. Both circuits have a number of qubits $n$ and a depth $d$ which can be varied independently, and we use $|\mathcal{C}| = 4$ (random) instances of each circuit for a given $n, d$. In experiments on quantum computers, we choose $n \in \{3, 5\}$ and $d \subseteq \{1, 3, 5, 7, 9\}$. Additionally, for a specific device (IBMQ Kolkata), we also perform a larger experiment with $n=12$ qubits and $d \in \{1, 5, 9\}$. The number of one- and two-qubit gates for a given $n, d$ depend on the circuit type and is discussed for each circuit type below. As discussed in Sec.~\ref{subsubsec:noisy-simulators} we repeat each of these experiments on noisy quantum computer simulators for comparison.

\subsubsection{Randomized benchmarking}

We use randomized benchmarking (RB) circuits~\cite{magesan3639robust, magesan2012characterizing, corcoles2013process, gambetta2012characterization, mckay2019three} as one benchmark problem in our experiments. An $n$-qubit, depth $d$ RB circuit is a sequence of $d$ random Clifford group elements $U_d  U_{d-1}  \cdots  U_1$, followed by a (classically computed) inverse element $U_{\rm inv} = (U_d  U_{d-1} \cdots  U_1)^{-1}$, such that the full circuit
\begin{equation}
    C = U_{\rm inv} U_d  U_{d-1} \cdots U_1
\end{equation}
is the identity operation (without noise). As such, the only bitstring that should be measured is $z_C = 0^n$ and we take the observable to be $\hat{A} = |z_C \rangle \langle z_C| = |0\rangle\langle 0|^{\otimes n}$.

In all experiments, we use a line of qubits and apply $2$-qubit RB sequences to each neighboring pair of qubits on the line. If the total number of qubits is odd we also apply a $1$-qubit RB sequence to the last qubit. The rationale for this choice is that a linear topology can be easily embedded in the connectivity graph of all quantum computers we consider in this work, and therefore this choice makes our benchmarks more consistent across different computers. 

Note that the parameter $d$ is the number of (parallel) random Clifford elements, not the actual physical depth of the circuit. Each Clifford element must be decomposed into two-qubit gates and single-qubit gates. The total number of two-qubit gates depends on both $d$ and the number of qubits $n$. In Table~\ref{tab:gate-count-rb}, we report the average number of two-qubit and single-qubit gates used in our experiments.
\begin{table}[htpb!]
    \centering
    \begin{tabular}{c|c|c|c}
        d  &   $n=3$                &  $n=5$                & $n=12$                \\ \hline
        1  &  3 (22)             & 7 (39)             & 19 (99)            \\
        3  &  6 (44)             & 11 (73)            & 36 (204)           \\
        5  &  9 (64)             & 18 (118)           & 53 (307)           \\
        7  &  12 (80)            & 24 (150)           & 73 (403)           \\
        9  &  15 (108)           & 31 (194)           & 89 (506)           \\
        12 &  18 (135)           & 37 (242)           & 115 (651)          \\
    \end{tabular}
    \caption{Average number of two-qubit (single-qubit) gates for an $n$-qubit, depth $d$ RB circuit. The average is taken over ten random instances. Note that the number of single-qubit gates may differ on different hardware due to the final compilation into native gates, but the number of two-qubit gates is hardware-independent.}
    \label{tab:gate-count-rb}
\end{table}
\subsubsection{Mirror circuits}

We also use mirror circuits~\cite{proctor2021scalable, proctor2022measuring} as a benchmark problem. Mirror circuits are similar to RB circuits in that they have a structure of random layers, but their final state $|z_C\rangle$ is randomized. This configuration allows a more uniform sampling of measurement errors.

An $n$-qubit, depth $d$ mirror circuit $C$ is a randomized sequence of $d$ Clifford layers and Pauli layers, where Clifford layers are organized in such a way to conjugate a Pauli layer into a rotated Pauli layer (see Fig. 1 of \cite{proctor2021scalable}). The full mirror circuit $C$ is equivalent to a random Pauli operator $\mathcal P$, thus the final ideal (noiseless) state is a random computational basis state $|z_C\rangle = \mathcal P|00\dots 0\rangle$, and we take the observable to be $\hat{A} = |z_C \rangle \langle z_C|$.


It is important to notice that the Clifford depth $d$ reported in our results for mirror circuits is the number of random Clifford layers. Since each Clifford layer is compiled into elementary gates, and since additional random Pauli layers are present in the circuit, the Clifford depth $d$ is different from the number of physical gates applied in the circuit. 
The average number of two-qubit and single-qubit gates used in our experiments are reported in Table \ref{tab:gate-count-mirror} for different values of $d$ and $n$.

\begin{table}[htpb!]
    \centering
    \begin{tabular}{c|c|c|c}
        d  & $n=2$          &  $n=5$            & $n=12$         \\ \hline 
        1  &  2 (26)       & 4 (41)           & 10 (95)     \\
        3  &  6 (46)       & 12 (68)          & 30 (158)    \\
        5  &  10 (68)      & 20 (98)          & 51 (223)    \\
        7  &  14 (87)      & 28 (125)         & 72 (284)    \\
        9  &  18 (108)     & 36 (154)         & 92 (350)    \\
        12 &  24 (138)     & 48 (197)         & 121 (448)   \\
    \end{tabular}
    \caption{Average number of two-qubit (single-qubit) gates for an $n$-qubit, depth $d$ mirror circuit. The average is taken over ten random instances. Note that the number of single-qubit gates may differ on different hardware due to the final compilation into native gates, but the number of two-qubit gates is hardware-independent.}
    \label{tab:gate-count-mirror}
\end{table}

\subsection{Quantum computers} \label{subsec:computers}

We test error mitigation techniques with each benchmark circuit on four quantum computers --- IBMQ Kolkata, IBMQ Lima, Rigetti Aspen-M2, and IonQ Harmony --- shown in~\figref{fig:methods-overview} and described in the following sections. We also perform experiments on noisy quantum computer simulators for comparison to hardware and for additional experiments. The noise models we use are described in Sec.~\ref{subsubsec:noisy-simulators}.


\subsubsection{IBMQ Lima} \label{subsubsec:ibmq-lima}

The IBMQ Lima computer consists of five superconducting transmon qubits arranged in a ``T-shape'' topology shown in~\figref{fig:methods-overview}(c). The error rates for the computer are listed in Table~\ref{tab:ibm-lima-device-specs} in Appendix~\ref{appx:quantum-computing-device-platforms}. In our $n = 3$ qubit experiments, we use the qubits with the lowest two-qubit error rates, namely the qubits labeled $(0, 1, 2)$. For $n = 5$ qubit experiments we use all qubits on the device.

\subsubsection{Rigetti Aspen-M2} \label{subsubsec:rigetti-aspen-m2}

The Rigetti Aspen-M2 computer consists of $80$ superconducting qubits arranged in a hexagonal lattice shown in~\figref{fig:methods-overview}(c). The error rates for the computer are listed in Table~\ref{tab:rigetti-aspen-m2-device-specs} in Appendix~\ref{appx:quantum-computing-device-platforms}. We perform $n = 3$ qubit experiments on Rigetti Aspen-M2 using a line of qubits with relatively low two-qubit error rates, namely the qubits labeled $(10, 17, 113)$ highlighted in Fig.~\ref{fig:methods-overview}(c). Due to limited device availability, we were only able to perform $n = 3$ qubit experiments on this computer.

\subsubsection{IonQ Harmony} \label{subsubsec:ionq-harmony}

The IonQ Harmony computer consists of $11$ trapped ion qubits with all-to-all connectivity, shown in~\figref{fig:methods-overview}(c). Unlike the IBMQ Lima and Rigetti Aspen-M2 computers, at the time of performing experiments on IonQ Harmony, it was not possible to select which qubits to use when submitting jobs or to check which qubits were used after jobs were completed. The average one-qubit and two-qubit gate errors are respectively $\epsilon_{1Q} = 0.0029$ and $\epsilon_{2Q} = 0.0073$. 

It is also worthwhile to note that, at the time of performing experiments, it was not possible (from the AWS platform) to disable compilation on IonQ Harmony, unlike on IBMQ Lima and on Rigetti Aspen-M2. Disabling compilation is important in error mitigation because techniques often insert gates that are logically trivial (e.g., $G G^\dagger$ in zero-noise extrapolation) or perform other modifications to produce circuits that are meant to be run exactly as specified. To avoid these problems when running ZNE on IonQ Harmony, we add barriers of single-qubit infinitesimal rotations as described in Appendix~\ref{sec:rotation-barriers-trick-ionq}. Due to limited device availability, we were only able to perform $n = 3$ qubit experiments on this computer.

\subsubsection{IBMQ Kolkata} \label{subsubsec:ibmq-kolkata}

To assess the performance of error mitigation on larger benchmark problem sizes (namely, $n = 12$ qubit experiments), we use the IBMQ Kolkata device based on the $27$-qubit superconducting chip  --- see Appendix~\ref{appx:quantum-computing-device-platforms} for the coupling map (\figref{fig:fake-kolkata-coupling-map}) and error rates (Table~\ref{tab:ibm-kolkata-device-specs}).

\subsubsection{Noisy quantum computer simulators} \label{subsubsec:noisy-simulators}

In addition to quantum computers, we also perform experiments on noisy quantum computer simulators (hereafter ``noisy simulators''). There are two primary reasons for this. First, this allows us to compare how error mitigation performs with simple noise models relative to actual quantum computers. Second, using noisy simulators allows us to circumvent practical limitations like device availability to perform additional experiments.

We consider two classes of noisy simulators: (i) one which implements a simple noise model of single-qubit depolarizing noise after each gate, and (ii) one which is based on the error rates  of a particular computer and so is meant to closely emulate that computer. The simple noise model we use is $1\%$ depolarizing noise after each two-qubit gate --- i.e., each two-qubit gate (CNOT) is replaced by~\eqref{eq:cnot_noise_model} with $p = 0.01$. In addition to this simple noise model, we also use a noise model based on the error rates of the IBMQ Lima computer (referred to as FakeLima). This noisy simulator has the same topology as IBMQ Lima shown in~\figref{fig:methods-overview}(c) and includes inhomogeneous single-qubit gate errors, two-qubit gate errors, and measurement errors based on the device characteristics in Table~\ref{tab:ibm-lima-device-specs}. Similarly, we used
IBMQ FakeKolkataV2 to classically simulate the hardware experiments performed on the real IBMQ Kolkata device. Its coupling map and error rates are reported in \figref{fig:fake-kolkata-coupling-map} and Table~\ref{tab:ibm-kolkata-device-specs}, respectively.

\section{Results}\label{sec:results}

As described in Sec.~\ref{sec:methods}, we applied ZNE(L), ZNE(R), and PEC to RB circuit and mirror circuit benchmarks on IBM, IonQ, and Rigetti quantum computers, as well as noisy quantum computer simulators. For each of these experiments, we compute the improvement factor~\eqref{eq:if} to quantify the performance of each error mitigation technique.

\begin{figure}
    \includegraphics[width=\columnwidth]{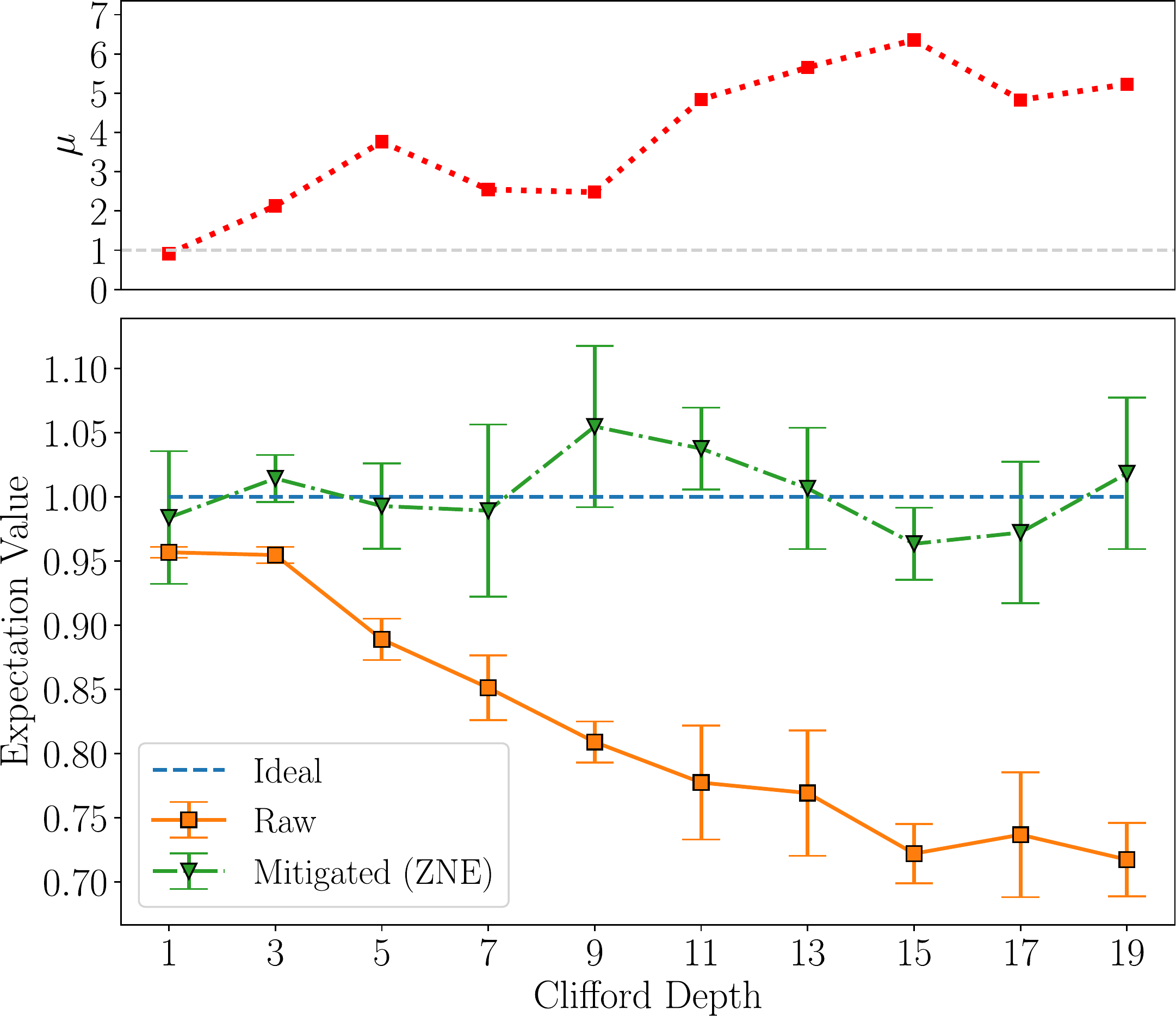}
    \caption{(Bottom panel) Unmitigated expectation values (orange squares) and the corresponding mitigated expectation values using ZNE(R) (green triangles) for $n = 3$ qubit RB circuits executed on IBMQ Lima. For all depths, the ideal expectation value is equal to 1 (dotted line). (Top panel) The improvement factor~\eqref{eq:if} at each depth for the results in the bottom panel.}
    \label{fig:mitigated-vs-unmitigated-results-bis}
\end{figure}

An example of the results from one particular experiment is shown in Fig.~\ref{fig:mitigated-vs-unmitigated-results-bis}. Here, we show the result from applying ZNE(R) to $n = 3$ qubit RB circuits of various depths on IBMQ Lima. At each depth, we generate $|\mathcal{C}| = 4$ RB circuits and evaluate the expectation value with and without error mitigation using $t = 1$ trial, and use this to compute the improvement factor~\eqref{eq:if}. As shown in Fig.~\ref{fig:mitigated-vs-unmitigated-results-bis}, the ``raw'' (unmitigated) results diverge from the ideal (noiseless) expectation value as the depth $d$ increases, while the ZNE(R) results are closer to the ideal expectation value but generally have higher variance. This is quantified in the improvement factor which here ranges from $\mu_{\rm ZNE(R)} \simeq 1$ to $\mu_{\rm ZNE(R)} \simeq 6$. In particular, all depths $d > 1$ show an improvement factor $\mu_{\rm ZNE(R)} > 1$, indicating ZNE(R) was always beneficial to use in this example.

\begin{figure*}
    \centering    
    \includegraphics[width=2.15\columnwidth]{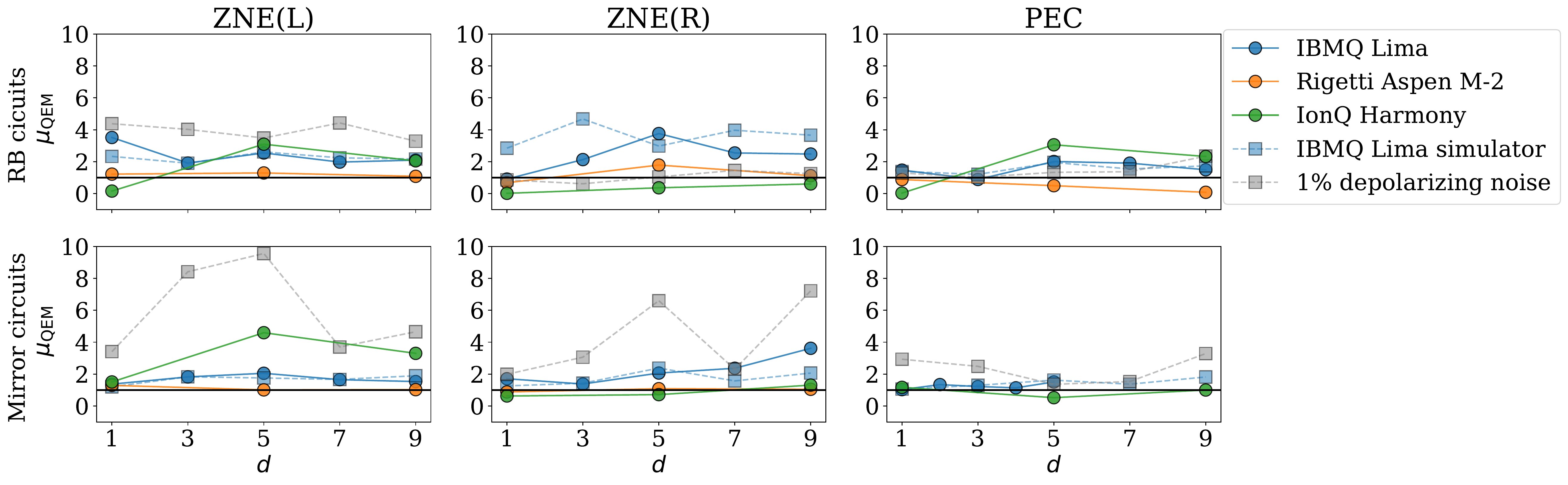}
    \caption{Improvement factor~\eqref{eq:if} results for $n = 3$ qubit experiments. From left to right, the quantum error mitigation techniques are ZNE with linear extrapolation, ZNE with Richardson extrapolation, and probabilistic error cancellation. The top panel shows improvement factors for  over $|\mathcal{C}| = 4$ randomized benchmarking circuits, and the bottom panel shows results for $|\mathcal{C}| = 4$ mirror circuits. Circle markers show quantum computer results and square markers show noisy quantum computer simulator results. In most cases, improvement factors are highest for the $1\%$ depolarizing noise model (grey squares), which is expected as this is the simplest noise model. Improvement factors on IBMQ Lima (blue circles) are almost always above $\mu = 1$, and the IBMQ Lima simulator results (blue squares) follow the computer results fairly closely. Improvement factors on Rigetti Aspen-M2 (orange circles) and IonQ Harmony (green circles) are frequently below $\mu = 1$ --- notably for ZNE(R) RB circuits --- indicating that error mitigation did not help in these experiments. Improvement factors from PEC are notably smaller than the ZNE experiments but are mostly above $\mu = 1$. }
    \label{fig:hardware-improvement-factor-plot}
\end{figure*}

We show the results of all $n = 3$ experiments in Fig.~\ref{fig:hardware-improvement-factor-plot}. Here, results are arranged in a grid displaying error mitigation techniques and benchmark problems, and different colored markers in each subplot show results on different quantum computers, including noisy simulators. As a baseline for comparison, we consider a very simple noise model of $1\%$ depolarizing noise (see Sec.~\ref{subsubsec:noisy-simulators}), and we find as expected that this simple noise model generally produces the largest improvement factors in experiments. (The interesting exception is ZNE(R) for which IBMQ Lima shows the largest improvement factors.)  On real quantum computers, there are additional sources of error including state preparation and measurement (SPAM) error as well as more complicated (in)coherent gate and crosstalk errors, so it is expected --- as we see in the results --- that these improvement factors are lower. However the improvement factors on the IBMQ Lima computer --- as well as the improvement factors on the IBMQ Lima simulator which follow the real computer fairly closely --- are still above $\mu = 1$, generally ranging between $\mu \simeq 1$ and $\mu \simeq 4$, indicating that error mitigation is beneficial on this device. 

The improvement factors for PEC are lower, between $\mu \simeq 1$ and $\mu \simeq 2$, so PEC was generally less beneficial to run than ZNE, but still more beneficial than no error mitigation for most backends (IBM, IonQ, and all simulators.). Moreover, we should take into account that PEC was applied assuming a very simplified noise model (depolarizing) and so we expect better performances when PEC is based on a more faithful noise characterization.
On IonQ Harmony, there are several cases where $\mu < 1$, especially in ZNE(R) experiments, so ZNE(R) was generally worse to use than no error mitigation on this computer, while ZNE(L) was more beneficial than no error mitigation. Interestingly, most improvement factors on Rigetti Aspen-M2 are close to $\mu = 1$, so error-mitigated results were generally the same as unmitigated results on this computer. Recall that our improvement factor~\eqref{eq:if} normalizes by additional shots.

\begin{figure*}
    \centering
    \includegraphics[width=2.15\columnwidth]{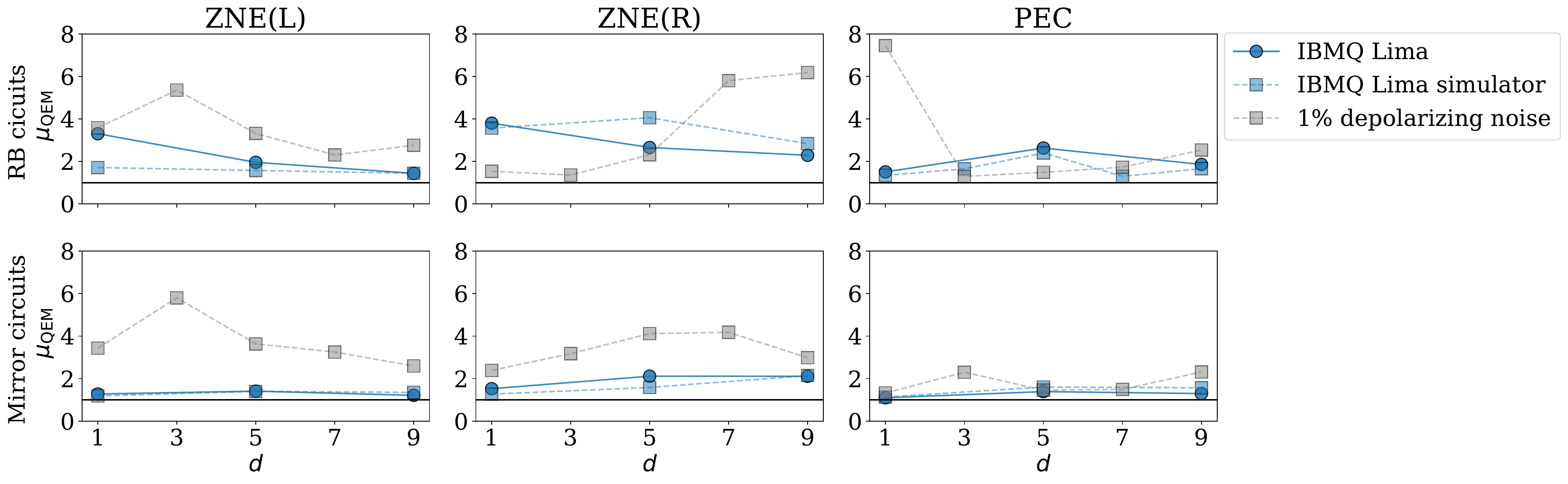}
    \caption{Improvement factor~\eqref{eq:if} results for $n = 5$ qubit experiments, in the same format as Fig.~\ref{fig:hardware-improvement-factor-plot}. In this case, we still see that improvement factors are usually highest on the $1\%$ depolarizing noise simulator, as expected. The average improvement factors for ZNE are comparable but slightly lower than those of the $n = 3$ qubit experiments, whereas the average improvement factors for PEC are noticeably larger than those of the $n = 3$ qubit experiments.}
    \label{fig:software-improvement-factor-plot}
\end{figure*}

We repeat the same type of experiments using $n = 5$ qubits and show these results in a similar format in Fig.~\ref{fig:software-improvement-factor-plot}. Here we were unable to perform experiments on Rigetti Aspen-M2 or IonQ Harmony due to limited device availability. We see again in these results that the improvement factors for the simple $1\%$ depolarizing noise model are generally the largest, as expected. The improvement factors on IBMQ Lima are comparable in magnitude to the $n = 3$ experiments, and in all cases, $\mu \ge 1$ so error mitigation was always beneficial. We also see that the IBMQ Lima simulator results follow the results of the real computer fairly closely, as in the $n = 3$ qubit experiment.


\begin{figure}
    \includegraphics[width=\columnwidth]{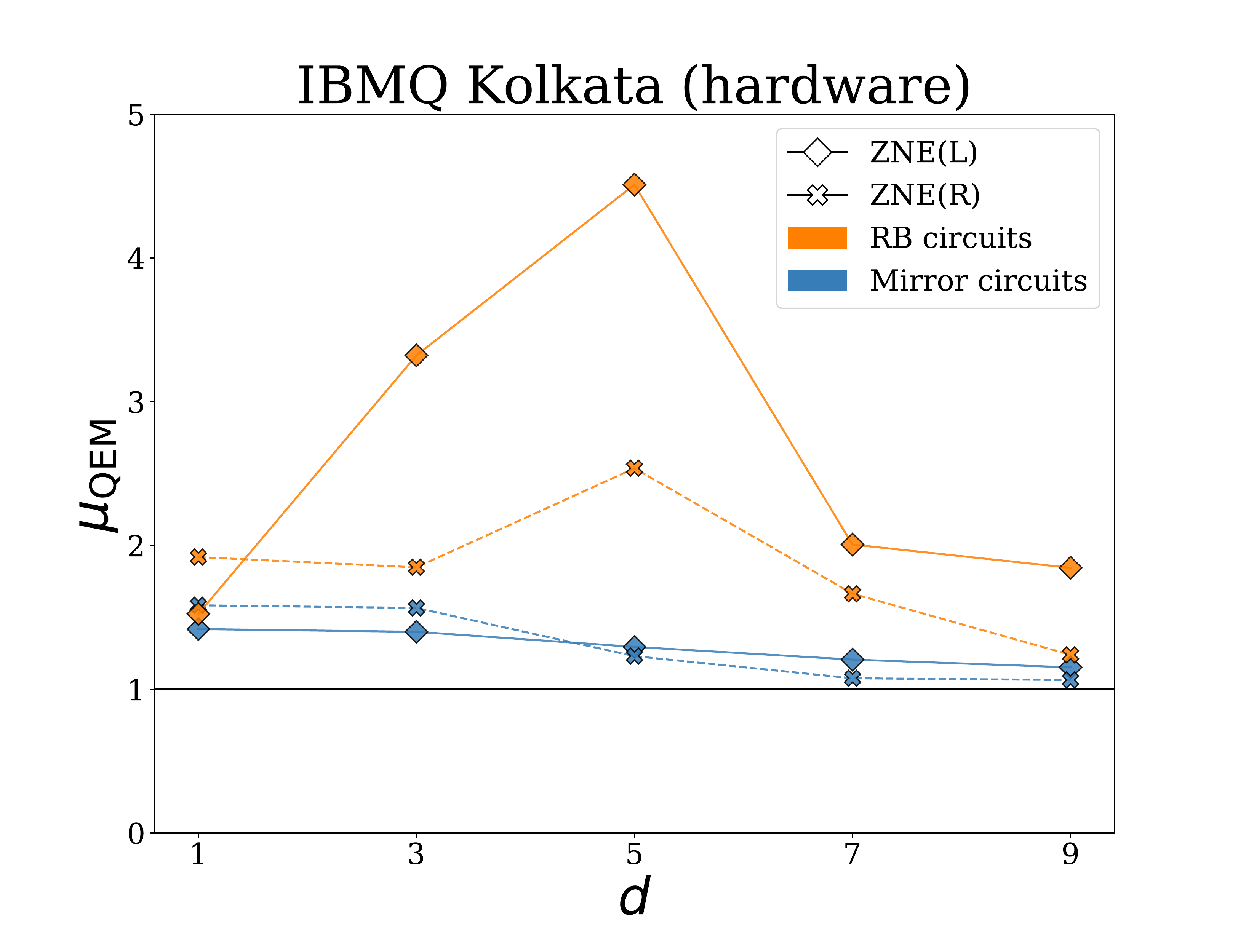}
    \\[\smallskipamount]
    \includegraphics[width=\columnwidth]{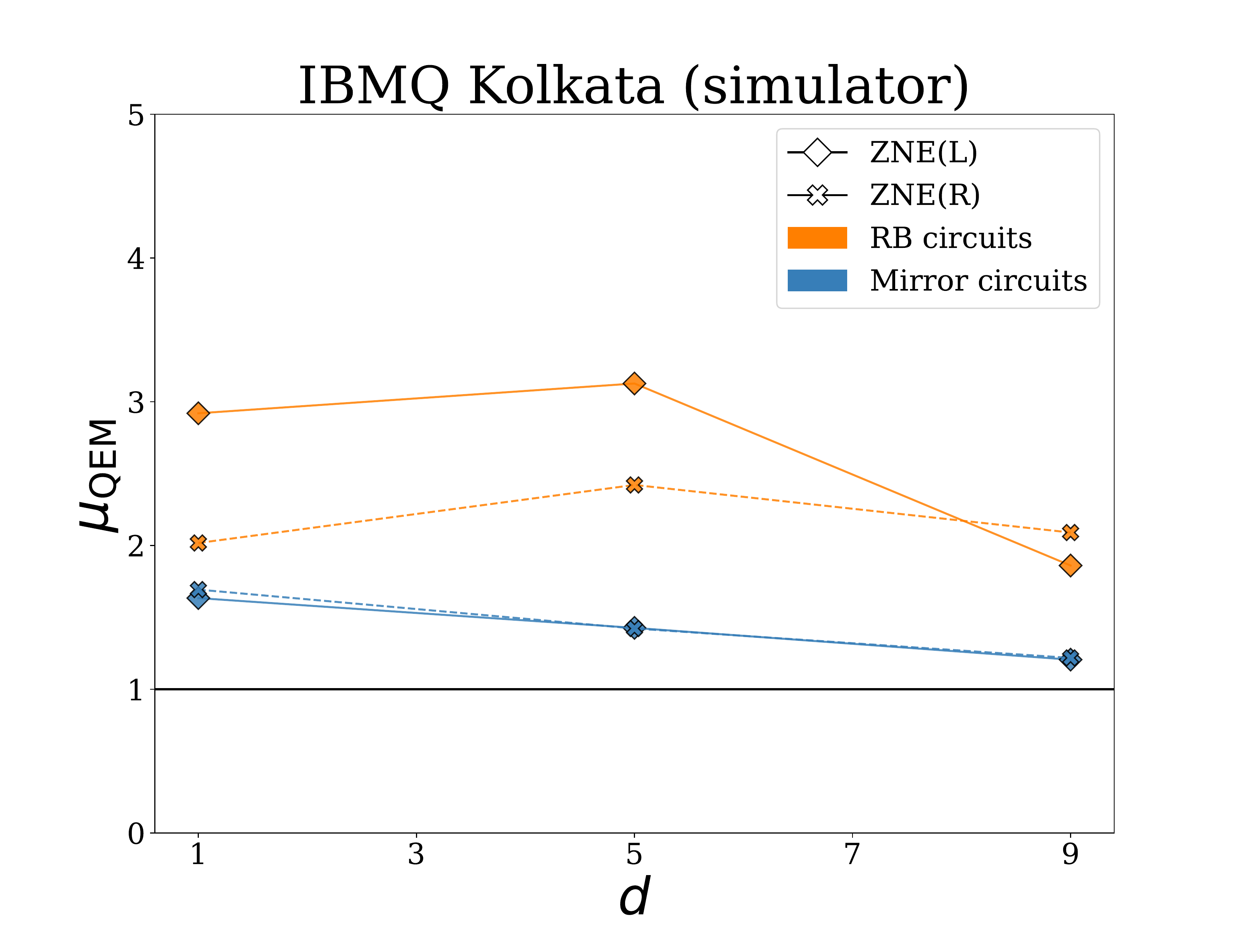}
    \\[\smallskipamount]
    \caption{Improvement factor~\eqref{eq:if} results for $n = 12$ qubit experiments on both a quantum device and a noisy simulator based on the IBMQ Kolkata computer (see Fig.~\ref{fig:fake-kolkata-coupling-map} for the coupling map and Table~\ref{tab:ibm-kolkata-device-specs} for error rates).}
    \label{fig:n12results}
\end{figure}

To further test the performance of quantum error mitigation as the problem size increases, we perform $n = 12$ qubit experiments on both a hardware device and a noisy quantum computer simulator and show these results in Fig.~\ref{fig:n12results}. The hardware device is the $27$-qubit IBMQ Kolkata computer and the noisy simulator is based on this Kolkata device (see Fig.~\ref{fig:fake-kolkata-coupling-map} for the coupling map and Table~\ref{tab:ibm-kolkata-device-specs} for the error rates). Here we see that improvement factors range from $\mu \simeq 1$ to $\mu \simeq 3$ indicating that error mitigation is still effective on larger problem sizes. The improvement factors for (parallel) randomized benchmarking circuits are higher than for mirror circuits in all cases, likely due to the fact that mirror circuits contain two-qubit gates across all edges. This is somewhat visible in $n = 3, 5$ qubit experiments but more accentuated here due to the larger problem size, and suggests that additional error mitigation on top of ZNE may be necessary to mitigate errors and crosstalk effects in larger applications. However, ZNE still performed better than the unmitigated experiments in all cases.

\section{Discussion}\label{sec:discussion}

\subsection{Positive features of our work}

A positive aspect of our experiments is that we apply error mitigation techniques ``out-of-the-box'', i.e., we do not tailor any techniques to the benchmark problems or computers we consider. Indeed, all experiments were performed with high-level API calls to quantum error mitigation software~\cite{larose2020mitiq} (see Appendix~\ref{app:implementation-details} for more on the implementation details). While tailoring techniques to specific experiments are likely to provide better results and is advisable in most applications, our approach gives a picture of what can be expected from quantum error mitigation in general applications. 

Another positive feature of our work is the comparison of results across several computers. This experimentally verifies that error mitigation can indeed be viewed as an algorithmic or software method independent of hardware, but our results also emphasize that the performance of error mitigation depends on the underlying computer. A clear example that illustrates this is zero-noise extrapolation with very deep circuits such that the final state is approximately the maximally mixed state. In this scenario scaling noise further does not produce any signal from which one can extrapolate, so ZNE has no advantage relative to no error mitigation. On a more accurate quantum computer, however, the final state may not be maximally mixed and noise may be able to be scaled with small-scale factors. On the opposite limit, if a quantum computer is already very accurate, the improvement factor due to error mitigation is necessarily small. In fact, in the limit of very weak noise, the bias of expectation values is negligible compared to the statistical variance which is typically not reduced by error mitigation (actually it is often increased by it).

Beyond experimental results, our work introduces a quantitative, problem-independent, and resource-normalized measure of the improvement of quantum error mitigation, the improvement factor (namely, \eqref{eq:if}). This is a natural and empirically-motivated measure that introduces a standardized metric for measuring and comparing error-mitigated quantum computer performance. For this reason, we expect the improvement factor metric to be used in other future experiments, beyond this work. We have shown the relation of our metric to the metric in~\cite{cirstoiu2022volumetric} so that results can be compared, and we encourage the use of quantitative metrics in future error mitigation work to continue this effort. Finally, we incorporated the notion of normalizing by additional resources (namely, shots) in our definition of the improvement factor, and experimentally showed error mitigation can still be beneficial even when adjusting by these extra resource requirements.

\subsection{Limitations of our work} \label{subsec:limitations-of-our-work}

While the improvement factor defined in~\eqref{eq:if} normalizes by additional shots used in quantum error mitigation, it does not normalize by additional qubits, gates, or circuits. While the error mitigation techniques we used in this work do not increase the number of qubits in executed circuits, other techniques do require more qubits in executed circuits, and accounting for this resource is important to understand the value of quantum error mitigation. 

In more detail, the techniques we use can increase both the number of gates and different circuits executed in an experiment. In particular, in ZNE the number of gates in each circuit is primarily increased but the number of circuits is only slightly increased, while in PEC the number of gates in each circuit is essentially the same as the original circuit but the number of circuits to execute is significantly higher. The optimal way to account for these additional resources is not immediately obvious, for example, whether to count additional circuits separately or to only count the total number of gates in all circuits. Similarly, even if additional circuits use the same number of qubits as the original circuit, they could be executed simultaneously by using additional qubits, so there is some subtlety concerning the most appropriate way to normalize for these space-time tradeoffs. Ultimately these questions are likely to depend on the particular computer being used, but the variety and flexibility of available error-mitigating techniques should be encouraging to builders and users of the many different quantum computer architectures being proposed.

Another potential resource in quantum error mitigation is pre-processing / noise characterization. For example, in PEC one needs to determine the noisy basis of the computer, and in measurement error mitigation one needs to characterize the confusion matrix. However, these examples and others usually do not grow with the size of the problem. Thus they are likely less important to account for in the improvement factor. (Recall that we assumed a particular noise model for PEC and did not perform gate characterization, so pre-processing cost is not present in our improvement factor results.) Applying ``out-of-the-box'' error mitigation techniques at the gate level may be paired with tailored and more complex noise models going beyond error calibration information produced by hardware providers \cite{schultz2022reducing,berg2022probabilistic}.   

Although we experimentally evaluate quantum error mitigation more generally than current literature (Table~\ref{tab:qem-experiments}), we still considered just two error mitigation techniques out of (roughly) dozens proposed in the literature. Additionally, both benchmark problems we used are based on random circuits --- while we expect our results to extend to structured circuits (say for time evolution), this needs to be experimentally verified. Due to limited device availability (i.e., which computers and how much computer time we had access to), we were only able to perform experiments on up to $n = 12$ qubits. This is fairly typical for error mitigation experiments (Table~\ref{tab:qem-experiments}) and we used noisy simulators to test the performance of error mitigation on larger problem sizes, but experiments on larger computers are still desirable. Last, we only considered experiments with a single error mitigation technique. Experiments composing two or more error mitigation techniques, for example, zero-noise extrapolation with dynamical decoupling and measurement error mitigation, will likely yield the largest improvement in applications. We leave quantifying this improvement (again normalized by additional resources used) and other points mentioned here for future work.

\subsection{Relationship to literature}

Our work adds a significant number of quantum error mitigation experiments relative the current literature (Table~\ref{tab:qem-experiments}). Notably, our work introduces a quantitative, problem-independent, and resource-normalized measure of the improvement of quantum error mitigation, and we compute this quantity on multiple quantum computers. With the notable exception~\cite{cirstoiu2022volumetric}, our work goes significantly beyond most experimental studies of quantum error mitigation which typically consider one technique for a specific experiment and do not explicitly evaluate a quantitative improvement of quantum error mitigation.

\subsection{Future outlook of QEM}

Based on our results and other experiments in the literature using error mitigation, we expect quantum error mitigation to be an essential component of virtually all experiments on ``NISQ'' computers~\cite{preskill2018quantum}, where we can take ``NISQ'' to roughly mean computers with up to $n \sim 10^2$ qubits capable of implementing $d \sim 10^3$ two-qubit gates. Indeed, depending on how much overhead is required for error correction, error mitigation techniques may continue to be important at even larger scales. 

Abstractly, quantum error mitigation can be viewed as combining NISQ computers with classical processors, and we have shown that this combination is still beneficial even when normalized by additional resources used. This combination is likely to be most beneficial when applied to problems that are classically hard~\cite{arute2019quantum}. In this setting, a quantum computer is used to get a rough solution to a hard problem, and a classical computer is used to improve the accuracy of the solution. We expect that combining the strengths of both devices in this manner will be necessary for solving problems too hard for either device to solve individually.

Furthermore, it is likely that quantum error mitigation will play an important role beyond NISQ computations, namely in error-corrected computations. Although QEM and QEC are sometimes thought of as separate techniques due to their different resource requirements and generality, they are similar in that both are algorithmic or software techniques to deal with errors in quantum computers. For example, dynamical decoupling --- largely considered to be a quantum error mitigation technique in many settings --- has been an important technique in recent quantum error correction experiments~\cite{acharya2022suppressing} to improve the fidelity of data qubits while syndrome measurements are performed. Further, Ref.~\cite{suzuki2022quantum} provides a more theoretical discussion about the application of quantum error mitigation in fault-tolerant quantum computing. We anticipate additional work connecting error mitigation to error correction at both the theoretical and experimental levels. Our results and experimental framework for obtaining these results~\cite{larose2020mitiq} provide some foundation for progress in this direction.

\section{Conclusion}

In this work we experimentally tested the performance of quantum error mitigation. Using an empirically-motivated metric that normalizes by the amount of additional resources used in a quantum error mitigation technique, we quantified the improvement of error mitigation using a variety of benchmark problems and quantum computers. In particular, we tested zero-noise extrapolation and probabilistic error mitigation on two benchmark problems and three quantum computers. The largest of such error mitigation benchmarks involved  quantum circuits acting on 12 qubits with more than 100 two-qubit gates and more than 600 single qubit gates. Our results show that error mitigation is on average more useful than no error mitigation, even when normalizing by the additional resources used and applying ``out-of-the-box'' error mitigation --- i.e., not tailoring the technique to the specific benchmark problem or the specific quantum computer. While these latter points are likely to provide further improvements and are encouraged in applications, our results provide a general picture of what can be expected of quantum error mitigation in practice.

Our definition of the improvement factor is, to our knowledge, the first quantitative metric to normalize by additional resources used in error mitigation, and we encourage the adoption of this or similar metrics in future work. It is also of interest to expand this metric for resources we do not account for here --- for example additional qubits and gates --- to more fully understand and quantify the value of quantum error mitigation in real experiments. This also can be considered with error mitigation techniques we did not use here --- for example dynamical decoupling)~\cite{viola1999dynamical, viola1998dynamical, zhang2014protected, pokharel2018demonstration, das2021adapt}, Clifford data regression~\cite{lowe2021unified, czarnik2021error}, and noise-extended probabilistic error cancellation~\cite{mari2021extending}. Additional experimental and theoretical results of this nature will help to further the progress made in this work and better understand the value of error mitigation in the larger context of quantum computing and quantum error correction.

\vspace{1em}
\textit{Note added:} While preparing our manuscript, we noticed a recent review of quantum error mitigation~\cite{cai2022quantum} which is similar in scope but discusses quantum error mitigation from a more theoretical rather than experimental perspective as in this paper.

\section*{Code and data availability}

Code and data are available upon reasonable HTTPS request to \href{https://github.com/unitaryfund/research/}{https://github.com/unitaryfund/research/}.



\section*{Acknowledgements}

We would like to thank Ethan Hickman for proposing, in a public discussion on the Mitiq Discord channel, the idea of using small rotations to block backend compilation. We would also like to thank Derek Wang for running the IBM Kolkata hardware experiments.
This work was supported by the U.S. Department of Energy, Office of Science, Office of Advanced Scientific Computing Research, Accelerated Research in Quantum Computing under Award Numbers DE-SC0020266 and DE-SC0020316 as well as by IBM under Sponsored Research Agreement No. W1975810. We thank IBM, Rigetti, and IonQ for providing access to their quantum computers. We thank the AWS Braket team for facilitating access to Rigetti and IonQ computers through the AWS Braket platform. The views expressed in this paper are those of the authors and do not reflect those of AWS, IBM, IonQ, or Rigetti.

\bibliographystyle{alpha}
\bibliography{refs}

\appendix

\section{Notation} \label{app:notation}

A summary of notation is shown in Table~\ref{tab:notation}.

\begin{table}
    \centering
    \begin{tabular}{|c|l|} \hline 
        $C$             & Quantum circuit (unitary) \\ \hline 
        $\rho$          & Final state of circuit $C$, $\rho = C |0 \rangle \langle 0 | C^\dagger$ \\ \hline 
        $\hat A$        & Hermitian observable \\ \hline
        $n$             & Number of qubits \\ \hline
        $d$             & Depth of circuit \\ \hline
        $N$             & Number of shots (samples) \\ \hline
        $A$             & Ideal expectation value ${\rm tr}[\rho \hat{A}]$ \\ \hline 
        $\mathcal E$    & Channel of a noisy computer (simulator) \\ \hline 
        $A'$            & Noisy expectation value ${\rm tr}[\mathcal{E} (\rho) \hat{A}]$ \\ \hline
        QEM             & A quantum error mitigation technique \\ \hline
        $A_{\rm QEM}$   & Error-mitigated expectation value \\ \hline
        $N_{\rm QEM}$    & Total number of shots used in the QEM technique \\ \hline
        $\mathcal C$    & A set of (benchmark) circuits \\ \hline
        $\mathcal{\hat{A}}$ & A set of (benchmark) observables \\ \hline
        $\mu_{\rm QEM}$  & Improvement factor~\eqref{eq:if} of QEM technique \\ \hline 
        ZNE             & Zero-noise extrapolation \\ \hline
        ZNE(L)          & ZNE with linear extrapolation \\ \hline
        ZNE(R)          & ZNE with Richardson extrapolation \\ \hline
        PEC             & Probabilistic error cancellation \\ \hline
    \end{tabular}
    \caption{Summary of notation. Note: The number of single-qubit and two-qubit gates in a circuit of depth $d$ depends on the circuit type --- see Sec.~\ref{subsec:benchmark-problems}.}
    \label{tab:notation}
\end{table}

\section{Implementation details} \label{app:implementation-details}

In this appendix, we discuss the implementation of the experiment and provide more details pertaining to how these experiments were programmatically carried out. 

\subsection{Experiment setup}\label{appx:experiment-setup}
For a given experimental run, the user specifies whether to run on a quantum hardware or simulator device, the platform to target (IBM, IonQ, or Rigetti) covered in~\ref{appx:quantum-computing-device-platforms}, the error mitigation method to apply (PEC or ZNE) covered in~\ref{appx:error-mitigation}, and the circuit type to consider (RB or mirror) covered in~\ref{appx:circuits}. 

Once we obtain either our RB or mirror circuit $C$, we define an executor function that takes the input circuit and returns an expectation value $\langle A \rangle$ where $A = \ket{z}\bra{z}$ for each circuit and where $z$ denotes the correct bitstring.

The circuits may contain gates that are not in the supported gatesets for the hardware device we are targeting to run on. In this case, we compile the gates in the circuit to gates in the supported gateset (more on this process in~\ref{appx:circuits}.) 

We then loop over each Clifford depth and within this loop, iterate over the number of trials we want to perform at each depth. For each iteration within our trial, we calculate the result of applying our error mitigation method. The resulting data is saved. Further information on the saved data can be found in~\ref{appx:software-and-experiment-data}.

\subsection{Software and experiment data}\label{appx:software-and-experiment-data}

Our experiments were carried out using Python 3.9. The error mitigation methods of PEC and ZNE were applied via version \texttt{0.18.0} of the Mitiq Python software package. The libraries Cirq (version \texttt{1.0.0}), \texttt{amazon-braket-sdk} (version \texttt{1.25.2}), and Qiskit (version \texttt{0.38.0}) were used to specify the circuits for our experiments. Further information on the Mitiq package can be found on the official GitHub repository \href{https://github.com/unitaryfund/mitiq}{https://github.com/unitaryfund/mitiq} as well as in~\cite{larose2020mitiq}.

The data obtained from our experiments and used to generate the plots in this work can be found in the \texttt{data} directory

\begin{center}
    \texttt{data/TYPE/QEM/CIRCUIT/PLATFORM/}
\end{center}

where $\texttt{TYPE} \in \{\texttt{hardware}, \texttt{software}\}$ describes whether the experiment was run on either an actual quantum device or a simulator, $\texttt{QEM} \in \{\texttt{pec}, \texttt{zne}\}$ describes the error mitigation method that was applied, $\texttt{CIRCUIT} \in \{\texttt{mirror}, \texttt{rb}\}$ describes the circuit type considered, and where $\texttt{PLATFORM} \in \{\texttt{ibmq}, \texttt{ionq}, \texttt{rigetti}, \texttt{depolarizing}\}$ describes on which platform the experimental data was obtained from.

Contained in each such directory is a subfolder with the following form
\begin{center}
    {\small \texttt{PLATFORM\_QEM\_CIRCUIT\_QUBITS\_MIN\_MAX\_SHOTS\_TRIALS}}
\end{center}

where \texttt{QUBITS} is the number of qubits used in the experiment, \texttt{MIN} is the minimum Clifford depth, \texttt{MAX} is the maximum Clifford depth, \texttt{SHOTS} is the total number of shots used in the experiment (this is 10,000 for all of our experiments) and \texttt{TRIALS} is the total number of trials carried out per experiment (this is 4 for all of our experiments).

In each such subfolder is a listing of files with the following prefixes:

\begin{itemize}
    \item \texttt{cnot\_counts}: The number of CNOT gates in the circuit.
    \item \texttt{noise\_scaled\_expectation\_values}: Noise-scaled expectation values (for ZNE only).
    \item \texttt{noisy\_values}: The non-scaled noisy expectation values (prior to applying error mitigation).
    \item \texttt{oneq\_counts}: The number of circuit instructions (modulo the number of CNOT operations).
    \item \texttt{true\_values}: The ideal values (these are always equal to 1).
    \item \texttt{mitigated\_values}: The error-mitigated values. 
\end{itemize}

Each row represents the value obtained at the depth corresponding to the index and each column represents the data obtained for a given trial.

Running the software in~\cite{unitary2022qem} that is responsible for capturing quantum device hardware experiment data requires possessing an AWS Braket account (for IonQ and Rigetti) an an IBM Quantum account (for IBM). Running the software on exclusively quantum simulators can be done without any such account access. 

To run the software on a simulator device the variable \texttt{use\_noisy\_simulator} should be set to \texttt{True} (and alternatively, \texttt{False} if the desire is to run on quantum device hardware). Setting the \texttt{mitigation\_type} variable to either \texttt{pec} and \texttt{zne} runs PEC or ZNE error mitigation, respectively. The type of circuit to use can be set via the variable \texttt{circuit\_type} to either \texttt{rb} for randomized benchmarking circuits or \texttt{mirror} for mirror circuits. Specifying the target platform can be done by setting the \texttt{hardware\_type} variable to either \texttt{ibmq}, \texttt{ionq}, or \texttt{rigetti} for IBM, IonQ, or Rigetti, respectively.

\subsection{Quantum computing device platforms}\label{appx:quantum-computing-device-platforms}

Access to the Rigetti Aspen-M2 and IonQ Harmony hardware devices were provided via the Amazon Braket service on AWS. Hardware information pertaining to qubit topology, error rates, etc. were obtained via the Amazon Braket API. Access to the IBM hardware Lima device and IBM FakeLima and FakeKolkataV2 simulator devices were provided by the IBM Quantum Compute Resources page~\cite{ibmq}.

\begin{table}[htpb!]
    \centering
    \begin{tabular}{c|c|c|c}
        Qubit & $\epsilon_{\rm 1Q}$ & $\epsilon_{\rm CX}$ & $\epsilon_{M}$  \\ \hline 
        0 & $9.028 \times 10^{-4}$ & $1.026 \times 10^{-2}$ & $2.740 \times 10^{-2}$ \\
        1 & $6.452 \times 10^{-4}$ & $1.083 \times 10^{-2}$ & $1.510 \times 10^{-2}$ \\
        2 & $6.016 \times 10^{-4}$ & $7.375 \times 10^{-3}$ & $1.700 \times 10^{-2}$ \\
        3 & $2.523 \times 10^{-4}$ & $1.570 \times 10^{-2}$ & $2.360 \times 10^{-2}$ \\
        4 & $6.167 \times 10^{-4}$ & $1.655 \times 10^{-2}$ & $4.410 \times 10^{-2}$
    \end{tabular}
    \caption{IBMQ Lima error rates for each qubit in the coupling map from~\figref{fig:methods-overview}(c). Here, $\epsilon_{\rm 1Q}$, $\epsilon_{\rm CX}$, and $\epsilon_{\rm M}$ represent the single-qubit $\sqrt{X}$-gate error, the average two-qubit CNOT error, and the readout assignment error, respectively~\cite{ibmq}.}
    \label{tab:ibm-lima-device-specs}
\end{table}

\begin{figure}
    \centering
    \includegraphics[width=0.8\columnwidth]{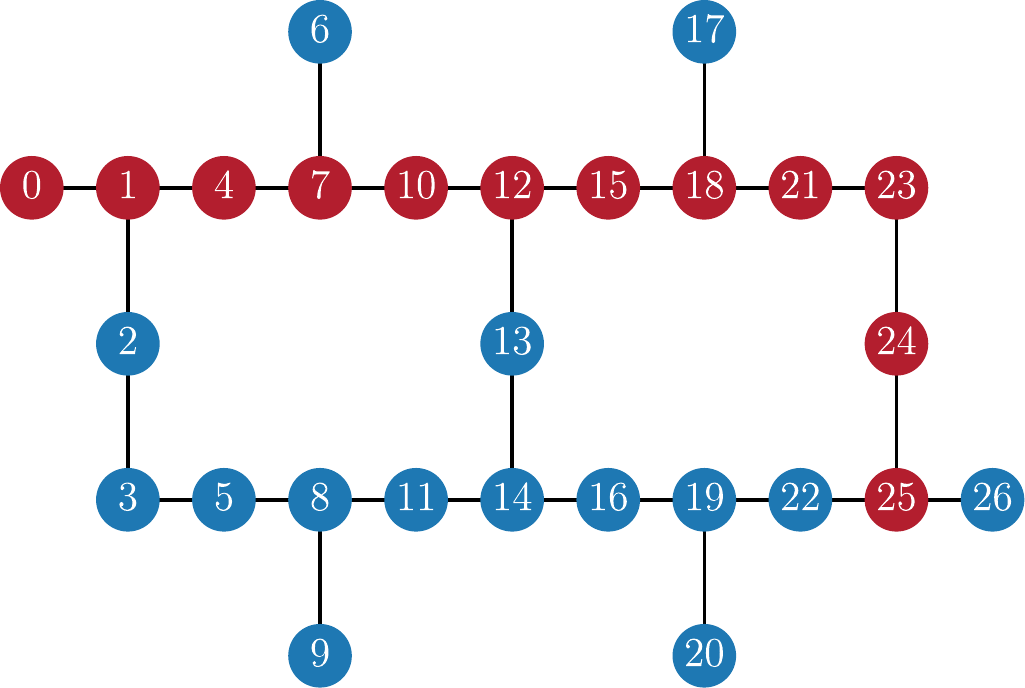}
    \caption{Coupling map for the $27$-qubit IBMQ Kolkata computer. We use this device (see Table~\ref{tab:ibm-kolkata-device-specs} for its properties) to perform $n = 12$ qubit error mitigation experiments on the red qubits. Results are shown in~\figref{fig:n12results}.}
    \label{fig:fake-kolkata-coupling-map}
\end{figure}
 
\begin{table}
    \centering
    \begin{tabular}{c|c|c|c}
        Qubit & $\epsilon_{\rm 1Q}$ & $\epsilon_{\rm CX}$ & $\epsilon_{M}$  \\ \hline 
        0 & $1.640 \times 10^{-4}$ & $3.997 \times 10^{-3}$ & $1.820 \times 10^{-2}$ \\
        1 & $1.339 \times 10^{-4}$ & $5.737 \times 10^{-}$ & $1.850 \times 10^{-2}$ \\
        4 & $1.660 \times 10^{-4}$ & $6.636 \times 10^{-3}$ & $2.750 \times 10^{-2}$ \\
        7 & $2.027 \times 10^{-4}$ & $1.259 \times 10^{-2}$ & $2.270 \times 10^{-2}$ \\
        10 & $4.948 \times 10^{-4}$ & $1.397 \times 10^{-2}$ & $1.320 \times 10^{-2}$ \\
        12 & $2.380 \times 10^{-4}$ & $9.326 \times 10^{-3}$ & $8.500 \times 10^{-3}$ \\
        15 & $2.626 \times 10^{-4}$ & $4.086\times 10^{-2}$ & $5.000 \times 10^{-3}$ \\        
        18 & $2.248 \times 10^{-4}$ & $1.052\times 10^{-2}$ & $6.200 \times 10^{-3}$ \\        
        21 & $1.772 \times 10^{-4}$ & $6.663\times 10^{-3}$ & $1.350 \times 10^{-2}$ \\        
        23 & $2.221 \times 10^{-4}$ & $5.340\times 10^{-3}$ & $8.100 \times 10^{-3}$ \\        
        24 & $2.858 \times 10^{-4}$ & $5.027\times 10^{-1}$ & $1.360 \times 10^{-2}$ \\        
        25 & $5.048 \times 10^{-4}$ & $3.368\times 10^{-1}$ & $6.600 \times 10^{-3}$       
    \end{tabular}
    \caption{IBMQ Kolkata error rates for each qubit used in our $n = 12$ qubit experiments (see Fig.~\ref{fig:fake-kolkata-coupling-map}). We use a noisy simulator based on these error rates to perform experiments. Here, $\epsilon_{\rm 1Q}$, $\epsilon_{\rm CX}$, and $\epsilon_{\rm M}$ represent the single-qubit $\sqrt{X}$-gate error, the average two-qubit CNOT error, and the readout assignment error, respectively~\cite{ibmq}.}
    \label{tab:ibm-kolkata-device-specs}
\end{table}

\begin{table}[htpb!]
    \begin{tabular}{ccc|ccc}
      & \textbf{Qubit specs} & & & \textbf{Edge specs} \\ \hline
      Qubit & Readout fidelity & & Edge & $\epsilon_{\rm CX}$ \\
      \hline
      10 & 99.3\% & & 10-17 & $3.79 \times 10^{-3}$ \\
      17 & 98.2\% & & 10-113 & $4.58 \times 10^{-3}$ \\
      113 & 94.7\% & \\
    \end{tabular}
  \caption{The Rigetti Aspen-M2 measures of the average two-qubit CNOT error ($\epsilon_{\rm CX}$) for the qubit edge and relative readout fidelity rates for the qubits in the device that we use in our experiments. Accessed from~\cite{aws2020braket}.}
    \label{tab:rigetti-aspen-m2-device-specs}
\end{table}

\subsection{Error mitigation}\label{appx:error-mitigation}

The PEC and ZNE error mitigation methods were applied using the Mitiq software package. 

\subsubsection{ZNE}\label{appx:zne}
The Mitiq package employs local folding~\cite{giurgica2020digital} and global folding~\cite{schultz2022reducing} as gate-level noise scaling methods for ZNE. In this work, we used global folding.

\begin{minipage}{\linewidth}
\begin{lstlisting}[label={listing:global-folding},caption={Apply global folding to a circuit in Mitiq.},captionpos=b,language=Python]
import qiskit
from mitiq.zne.scaling import fold_global

# Define a quantum circuit.
qubits = qiskit.QuantumRegister(2)
circuit = qiskit.QuantumCircuit(qubits)
circuit.h(0)
circuit.cnot(0, 1)

# Apply global folding.
scaled_circuit = fold_global(circuit, scale_factor=3)
\end{lstlisting}
\end{minipage}

\begin{figure}
    \centering
\scalebox{1.3}{
    \Qcircuit @C=.5em @R=0em @!R {
    & \gate{} & \multigate{1}{} & \qw & \qw & &  
    & \gate{} & \multigate{1}{} & \qw & \qw & \qw &  \multigate{1}{} & \gate{} & \qw & \gate{} & \multigate{1}{} & \qw & \qw \\ 
    & \qw & \ghost{} & \gate{} & \qw & \push{\rule{.5em}{0em}\rule{.5em}{0em}} & 
    &  \qw & \ghost{} & \gate{} & \qw & \gate{} & \ghost{} & \qw & \qw & \qw & \ghost{} & \gate{} & \qw  \\
    \\    
    &  & {\underbrace{\hspace{4em}}_{\text{$G$}}}  & & & & & & {\underbrace{\hspace{4em}}_{\text{$G$}}} & & & & {\underbrace{\hspace{4em}}_{\text{$G^{\dag}$}}} & & & & {\underbrace{\hspace{4em}}_{\text{$G$}}}  \\    
}}
    \caption{Example of the global folding technique applied to an arbitrary circuit. For a given collection of gates denoted by $G$, we increase the effective length of the overall circuit by creating $GG^{\dag}G$.}
    \label{fig:global-folding-example}
\end{figure}
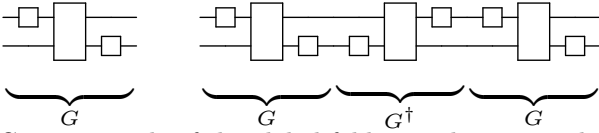

Global folding increases the effective length of the quantum circuit by compiling the input circuit with a larger number of gates. Each set of layers in the circuit is replaced by $G G^{\dag} G$. Since $G G^{\dag} = I$, in the case where we are running our circuit on an ideal simulator this has no effect on the circuit. However, in the case where one uses a noisy device, this increases the noise and effective gate errors of the computation. An arbitrary example that depicts the global folding technique is shown in~\figref{fig:global-folding-example}. 

An example of how to make use of global folding in Mitiq is provided in Listing~\ref{listing:global-folding} that makes use of $\lambda=3$ being a scale factor. Note that the smallest scale factor one can select is $\lambda=1$ that corresponds to not performing any folding where selecting $\lambda=3$ folds all of the gates in the circuit. For any scale factors $\lambda > 3$ in Mitiq, this folds some or all gates in the circuit.

For ZNE, we applied linear and Richardson extrapolation methods provided as factory objects in Mitiq as \texttt{LinearFactory} and \texttt{RichardsonFactory}, respectively. 

\begin{lstlisting}[label={listing:zne},language=Python, caption={Applying ZNE in Mitiq using Richardson extrapolation with noise scale factors $\lambda_i \in \{1,2,3\}$.},captionpos=b]
from mitiq import zne

zne_value = zne.execute_with_zne(
    circuit, 
    execute, 
    scale_noise=zne.scaling.fold_global, 
    factory=zne.inference.RichardsonFactory(scale_factors=[1, 2, 3])
)
\end{lstlisting}

\subsubsection{PEC}\label{appx:pec}

To apply probabilistic error cancellation~\cite{temme2017error, endo2018practical, zhang2020error} we use the associated Mitiq module. We extract the two-qubit error probability $p_{\rm 2Q}$ from the backend properties as reported by the hardware vendor. We use this information together with the utilities in \texttt{mitiq.representations} to generate the quasi-probability representations for all the two-qubit operations acting on neighboring qubits of the quantum processor. We store the result as a list (\texttt{representations}) of \texttt{mitiq.pec.OperationRepresentation} objects.
After this preliminary step, we can obtain all the error-mitigated expectation values as described in the next code block.

\begin{lstlisting}[label={listing:pec},language=Python,caption={Apply PEC using Mitiq.},captionpos=b,]
from mitiq import pec

pec_value = pec.execute_with_pec(
    circuit, 
    execute, 
    representations, 
    num_samples=num_samples,
    random_state=local_seed,
)
\end{lstlisting}

\subsection{Circuits}\label{appx:circuits}

In order to construct our RB circuits, we define an RB pattern by splitting the qubits into 2-qubit pairs. We then generate a generic RB sequence via the Qiskit library. If the circuit is to be run on either Rigetti or IonQ, we perform a conversion of the Qiskit circuit to a Braket circuit. 

\begin{lstlisting}[language=Python, label={listing:rb-circuits},caption={Defining RB circuits in Mitiq.},captionpos=b,]
import qiskit.ignis.verification.randomized_benchmarking as rb

def get_circuit(depth, seed):
    circuit = rb.randomized_benchmarking_seq(
        length_vector=[depth], 
        rb_pattern=rb_pattern,
        group_gates="0", 
        rand_seed=seed,
    )[0][0][0]
\end{lstlisting}

We generate mirror circuits via the \texttt{generate\_mirror\_circuit} function from Mitiq. Mirror circuits parameterize the number of random Clifford layers to be generated. 

\begin{lstlisting}[language=Python, label={listing:mirror-circuits},caption={Defining mirror circuits in Mitiq.},captionpos=b]
from mitiq.benchmarks import generate_mirror_circuit

def get_circuit(depth, seed):
    circuit, correct_bitstring = generate_mirror_circuit(
        nlayers=depth,
        two_qubit_gate_prob=1.0,
        connectivity_graph=computer,
        two_qubit_gate_name="CNOT",
        seed=seed,
        return_type=return_type,
    )
\end{lstlisting}

\subsubsection{Circuit compilation}\label{appx:circuit-compilation}
At the time of this writing, both Rigetti and IonQ hardware support the option of \emph{verbatim compilation}; a method that directs the compiler to run the specified circuit exactly as defined without adding any modifications. We attempt to disable automatic compilation by the platform service or QPU providers in order to have as much control as possible on the compiled circuit and hence error mitigation scaling.

The usage of verbatim compilation requires that every gate in the circuit is a gate that is natively supported by the hardware it is running on. For Aspen-M2, the native gate set is $\{\text{RX}, \text{RZ}, \text{CPHASE}, \text{CZ}, \text{XY}\}$. In our \texttt{compile\_to\_rigetti\_gateset} function, we iterate through every instruction in our circuit and compile all of the gates into equivalent ones that are in the supported native gateset. As IonQ only recently added support for verbatim compilation, we were not able to take advantage of this option for the IonQ Harmony experiments.

\subsubsection{Task batching}\label{appx:task-batching}

The AWS braket platform allows for task batching; the ability to launch jobs in parallel. For the majority of our experiments, serial execution within the allotted device time windows were sufficient to carry out a complete run of our experiment. One exception to this was the PEC experiments on Rigetti and IonQ hardware. In order to ensure these experiments ran within the allotted device time window, we needed to make use of the batching functionality. 

\subsection{Rotation barriers} \label{sec:rotation-barriers-trick-ionq}
Many hardware backends internally optimize circuits before actually running physical gates on a quantum processor. This can be a problem for some error mitigation techniques. For example, in  ZNE we want to run circuits whose depth is intentionally increased by unitary folding $G \rightarrow G G^\dag G$ (see Fig. \ref{fig:global-folding-example}). However, if the internal optimizer of a backend detects that $G^\dag G = I$, it will simplify the unitary folding structure such that any noise scaling effect is lost. This is a relevant practical issue for gate-level ZNE. 

The best way to avoid this effect is to optimize circuits before applying ZNE and to switch off any further circuit optimizations on the backend side. When this is not possible, a simple workaround is the addition of barriers of infinitesimal gates that generate a negligible unitary effect on the quantum state but that can block the action of circuit optimizers. 

In practice, in our experiments we apply ZNE by using a slightly modified version of the unitary folding rule:
\begin{equation}\label{eq:folding_with_barriers}
    G \rightarrow G R_1 G^\dagger R_2 G,
\end{equation}
where $G$ is the circuit acting on $n$ qubits and $R_j=[R_x(\epsilon_{x, j}) R_y(\epsilon_{y, j}) R_z(\epsilon_{z, j})]^{\otimes n}$ is the tensor-product of $n$ infinitesimal rotations. For each circuit block of the unitary folding structure, we apply a rotation barrier. The way in which the small rotation angles are chosen is quite arbitrary as long as they are sufficiently small but nonzero. In our experiments, we randomly sample between two fixed small angles ($\pm 10^{-4}$) as reported in the next code block.

\begin{lstlisting}[language=Python, label={listing:rotation-barriers},caption={Function returning a layer of infinitesimal rotations. Each layer is applied as a barrier between circuit blocks as shown in~\eqref{eq:folding_with_barriers}.},captionpos=b]
import cirq
import numpy as np 

def make_rotation_barrier(
    circuit,
    delta=0.0001,
):
    """Returns a barrier of infinitesimal rotations."""
    qubits = list(circuit.all_qubits())
    delta_x = np.random.choice([1.0, -1.0]) * delta
    delta_y = np.random.choice([1.0, -1.0]) * delta
    delta_z = np.random.choice([1.0, -1.0]) * delta
    barrier = cirq.Circuit()
    for q in qubits:
        barrier.append(cirq.rx(delta_x)(q))
        barrier.append(cirq.ry(delta_y)(q))
        barrier.append(cirq.rz(delta_z)(q))
    return barrier
\end{lstlisting}

Note that all our ZNE experiments on IonQ hardware have been done via the AWS cloud platform before {\it verbatim compilation} became available. Today {\it verbatim compilation} of quantum circuits into native gates is supported for both IonQ and Rigetti devices. Therefore, the workaround of applying rotation barriers is probably not necessary anymore to reproduce similar experiments.

\begin{table*}
    \centering
    \begin{tabular}{|c|c|c|c|c|c|c|}
        \hline
        Platform & Computer & QEM & Extrapolation & Circuit & Qubits & Simulator \\ \hline 
        IBM & Lima & ZNE & Richardson & RB & 3 & No \\
        IBM & Lima & ZNE & linear & RB & 3 & No \\        
        IBM & Lima & PEC & - & RB & 3 & No \\ \hline
        IBM & Lima & ZNE & Richardson & RB & 5 & No \\
        IBM & Lima & ZNE & linear & RB & 5 & No \\        
        IBM & Lima & PEC & - & RB & 5 & No \\ \hline 
        IBM & Lima & ZNE & Richardson & Mirror & 3 & No \\
        IBM & Lima & ZNE & linear & Mirror & 3 & No \\        
        IBM & Lima & PEC & - & Mirror & 3 & No \\ \hline                
        IBM & Lima & ZNE & Richardson & Mirror & 5 & No \\
        IBM & Lima & ZNE & linear & Mirror & 5 & No \\        
        IBM & Lima & PEC & - & Mirror & 5 & No \\ \hline
        IBM & Kolkata & ZNE & Richardson & RB & 3 & No \\
        IBM & Kolkata & ZNE & linear & RB & 3 & No \\ \hline
        IBM & Kolkata & ZNE & Richardson & Mirror & 3 & No \\
        IBM & Kolkata & ZNE & linear & Mirror & 3 & No \\ \hline        
        IonQ & Harmony & ZNE & Richardson & RB & 3 & No \\
        IonQ & Harmony & ZNE & linear & RB & 3 & No \\        
        IonQ & Harmony & PEC & - & RB & 3 & No \\ 
        \hline        
        IonQ & Harmony & ZNE & Richardson & Mirror & 3 & No \\
        IonQ & Harmony & ZNE & linear & Mirror & 3 & No \\        
        IonQ & Harmony & PEC & - & Mirror & 3 & No \\
        \hline                
        Rigetti & Aspen-M2 & ZNE & Richardson & RB & 3 & No \\
        Rigetti & Aspen-M2 & ZNE & linear & RB & 3 & No \\        
        Rigetti & Aspen-M2 & PEC & - & RB & 3 & No \\ \hline                
        Rigetti & Aspen-M2 & ZNE & Richardson & Mirror & 3 & No \\
        Rigetti & Aspen-M2 & ZNE & linear & Mirror & 3 & No \\\hline 
        IBM & FakeLima & ZNE & Richardson & RB & 3 & Yes \\
        IBM & FakeLima & ZNE & linear & RB & 3 & Yes \\        
        IBM & FakeLima & PEC & - & RB & 3 & Yes \\ \hline
        IBM & FakeLima & ZNE & Richardson & RB & 5 & Yes \\
        IBM & FakeLima & ZNE & linear & RB & 5 & Yes \\        
        IBM & FakeLima & PEC & - & RB & 5 & Yes \\ \hline 
        IBM & FakeLima & ZNE & Richardson & Mirror & 3 & Yes \\
        IBM & FakeLima & ZNE & linear & Mirror & 3 & Yes \\        
        IBM & FakeLima & PEC & - & Mirror & 3 & Yes \\ \hline                
        IBM & FakeLima & ZNE & Richardson & Mirror & 5 & Yes \\
        IBM & FakeLima & ZNE & linear & Mirror & 5 & Yes \\        
        IBM & FakeLima & PEC & - & Mirror & 5 & Yes \\ \hline        
        IBM & FakeKolkataV2 & ZNE & Richardson & RB & 12 & Yes \\
        IBM & FakeKolkataV2 & ZNE & linear & RB & 12 & Yes \\ \hline
        IBM & FakeKolkataV2 & ZNE & Richardson & Mirror & 12 & Yes \\
        IBM & FakeKolkataV2 & ZNE & linear & Mirror & 12 & Yes \\ \hline
        AWS & $1\%$ depolarizing noise & ZNE & Richardson & RB & 3 & Yes \\
        AWS & $1\%$ depolarizing noise & ZNE & linear & RB & 3 & Yes \\        
        AWS & $1\%$ depolarizing noise & PEC & - & RB & 3 & Yes \\ \hline
        AWS & $1\%$ depolarizing noise & ZNE & Richardson & RB & 5 & Yes \\
        AWS & $1\%$ depolarizing noise & ZNE & linear & RB & 5 & Yes \\        
        AWS & $1\%$ depolarizing noise & PEC & - & RB & 5 & Yes \\ \hline 
        AWS & $1\%$ depolarizing noise & ZNE & Richardson & Mirror & 3 & Yes \\
        AWS & $1\%$ depolarizing noise & ZNE & linear & Mirror & 3 & Yes \\        
        AWS & $1\%$ depolarizing noise & PEC & - & Mirror & 3 & Yes \\
        \hline
    \end{tabular}
    \caption{Summary of experiments performed.}
    \label{tab:experiment-details}
\end{table*}

\end{document}